\documentclass[prd,
superscriptaddress,
nofootinbib,%
tightenlines ]{revtex4}
\usepackage{epsfig}
\usepackage{amssymb}
\usepackage{bbold}

\newcommand{\ben}{\begin{displaymath}}
\newcommand{\een}{\end{displaymath}}
\newcommand{\be}{\begin{equation}}
\newcommand{\ee}{\end{equation}}
\newcommand{\bea}{\begin{eqnarray}}
\newcommand{\eea}{\end{eqnarray}}

    \newcommand{\fet}[1]{\mbox{\boldmath $#1$}}

\usepackage{color}

\def\Xint#1{\mathchoice
   {\XXint\displaystyle\textstyle{#1}}%
   {\XXint\textstyle\scriptstyle{#1}}%
   {\XXint\scriptstyle\scriptscriptstyle{#1}}%
   {\XXint\scriptscriptstyle\scriptscriptstyle{#1}}%
   \!\int}
\def\XXint#1#2#3{{\setbox0=\hbox{$#1{#2#3}{\int}$}
     \vcenter{\hbox{$#2#3$}}\kern-.5\wd0}}

\def\dashint{\Xint-}

\begin{document}
\title{Effective field theory for shallow P-wave states}
\author{E.~Epelbaum}
 \affiliation{Institut f\"ur Theoretische Physik II, Ruhr-Universit\"at Bochum,  D-44780 Bochum,
 Germany}
\author{J.~Gegelia}
 \affiliation{Institut f\"ur Theoretische Physik II, Ruhr-Universit\"at Bochum,  D-44780 Bochum,
 Germany}
\affiliation{Tbilisi State  University,  0186 Tbilisi,
 Georgia}
\author{H.~P.~Huesmann}
 \affiliation{Institut f\"ur Theoretische Physik II, Ruhr-Universit\"at Bochum,  D-44780 Bochum,
 Germany} 
\author{Ulf-G.~Mei{\ss}ner}
 \affiliation{Helmholtz Institut f\"ur Strahlen- und Kernphysik and Bethe
   Center for Theoretical Physics, Universit\"at Bonn, D-53115 Bonn, Germany}
 \affiliation{Institute for Advanced Simulation, Institut f\"ur Kernphysik
   and J\"ulich Center for Hadron Physics, Forschungszentrum J\"ulich, D-52425 J\"ulich,
Germany}
\affiliation{Tbilisi State  University,  0186 Tbilisi,
 Georgia}
\author{Xiu-Lei Ren}
\affiliation{Institut f\"ur Kernphysik \&  Cluster of Excellence PRISMA$^+$,
 Johannes Gutenberg-Universit\"at  Mainz,  D-55128 Mainz, Germany}

\begin{abstract}
We discuss the formulation of a non-relativistic effective field
theory for two-body  P-wave scattering in the presence of shallow
states and critically address various approaches to renormalization
proposed in the literature.  It is demonstrated that the consistent
renormalization involving only a finite number of parameters 
in the well-established formalism with auxiliary dimer fields corresponds to
the inclusion of an infinite number of counterterms in the
formulation with contact interactions only. We also discuss the
implications from the Wilsonian renormalization group analysis of
P-wave scattering. 
\end{abstract}

\maketitle

\section{Introduction}

In the early 1990s, Weinberg has argued that nuclear forces and
low-energy nuclear dynamics can be systematically analyzed using an
effective chiral Lagrangian~\cite{Weinberg:1990rz,Weinberg:1991um}. 
Today, 30 years after these seminal papers, chiral
effective field theory (EFT) has reached maturity to become
a precision tool in the two-nucleon sector
\cite{Epelbaum:2014sza,Entem:2017gor,Reinert:2017usi,Hernandez:2017mof,Filin:2019eoe,Reinert:2020mcu},
see Refs.~\cite{Epelbaum:2008ga,Machleidt:2011zz,Epelbaum:2019kcf,Tews:2020hgp,Piarulli:2020mop,Hammer:2019poc}
for review articles. In spite of this success, there is still no consensus on
what concerns the proper renormalization and power counting for few-body systems in chiral EFT. 
As any realistic quantum field theory (QFT), chiral EFT requires
regularization of ultraviolet (UV) divergences by means of some kind
of a regulator, say a cutoff. As the effective Lagrangian contains all terms allowed by
the underlying symmetries, it is, in principle, possible to completely absorb the
regulator (cutoff) dependence of physical quantities in a redefinition of
parameters entering the effective Lagrangian, provided the
applied regularization does not violate the underlying symmetries.
Since the effective Lagrangian contains an infinite number of terms,
one needs a systematic power counting scheme to
classify various terms in the Lagrangian according to their importance
and to set up an expansion of physical quantities in terms of the corresponding  
small parameter(s).  A word of caution  is in order here. It is a
common practice in QFTs to split the bare  parameters and fields into
renormalized ones that give rise to the renormalized  part of the Lagrangian
and the corresponding counterterms.  While in renormalizable  perturbative  
QFTs, all physical quantities are calculated within power-series expansions in
terms of renormalized coupling constants,  in chiral EFT the
expansion is performed in small momenta and masses.
This introduces an additional complication, since the relation
between  the expansion of the physical quantities in terms of small
parameters and the corresponding expansion of the effective
Lagrangian reflects the whole complexity of the QFT regularization and renormalization
and becomes particularly nontrivial for systems, whose description requires
performing certain kinds of nonperturbative resummations. First, one
needs to specify whether the power counting for the effective
Lagrangian is formulated in terms of bare or renormalized parameters. 
While this issue is irrelevant in the purely mesonic sector of chiral
EFT if one uses dimensional regularization (DR), things
start becoming more complicated already in the single-nucleon sector.
Using the heavy baryon approach \cite{Jenkins:1990jv,Bernard:1992qa} in combination with
DR allows one to deal with  this issue also for this
case.   However, starting from the two-nucleon sector, it seems impossible
to find a formulation that would allow one making no distinction
between the power counting being applied to the bare or the renormalized parameters.
In this context, it is important to keep in mind that  the numerical values
and, therefore, the relative importance of bare parameters depend on
the cutoff and are controlled by the Wilsonian renormalization group (RG)
equations \cite{Wilson:1974mb}, while the renormalized couplings depend on the
renormalization scales as dictated by the Gell-Mann and Low RG equations
\cite{GellMann:1954fq, Callan:1970yg,Symanzik:1970rt}. These two kinds of
RG equations are similar in spirit but  not identical. 

Our understanding of the chiral EFT approach for nuclear
systems proposed by Weinberg in Refs.~\cite{Weinberg:1990rz,Weinberg:1991um} is that the power
counting suggested in these works is supposed to be applied to the
renormalized Lagrangian, i.e., to the interaction terms with renormalized
parameters. In Ref.~\cite{Epelbaum:2017byx}, we have explicitly specified
the renormalization conditions corresponding to Weinberg's power counting
with all  renormalized LECs scaling according to naive dimensional
analysis (NDA) for  two-nucleon S-wave scattering in pionless EFT.
We believe that the frequently repeated claim of the
inconsistency of Weinberg's power counting, see, e.g., the recent
review article \cite{Hammer:2019poc}, stems from
interpreting it as the power
counting for the bare Lagrangian, see Ref.~\cite{Epelbaum:2017byx} for
a discussion.  We emphasize, however, that the implementation of the
scheme proposed in Ref.~\cite{Epelbaum:2017byx} in chiral EFT with pions treated as
dynamical degrees of freedom is plagued with severe technical
issues, see, however, Ref.~\cite{Kaplan:2019znu} for recent analytic calculations in
the chiral limit. Therefore, in practice, one usually utilizes a finite-cutoff
formulation of chiral EFT, where renormalization is carried out
implicitly by expressing the bare parameters in terms of observable
quantities, see Refs.~\cite{Epelbaum:2019kcf,Lepage:1997cs} for details. 

An alternative approach to formulating a systematic power counting 
via self-consistent renormalization conditions is provided by the 
Wilsonian RG method. Its application to the nucleon-nucleon (NN)
scattering problem in pionless EFT was pioneered in
Ref.~\cite{Birse:1998dk}, followed by numerous works addressing
various aspects of this formalism. 
While the Wilsonian RG and the associated power counting for the
nuclear forces have been extensively discussed in the literature,  see, e.g.,
Refs.~\cite{Birse:1998dk,Birse:2009my,Harada:2010ba,Harada:2006cw}, 
the term ``RG invariance'' is also being used in a different setting
as discussed e.g.~in Refs.~\cite{Hammer:2019poc,Valderrama:2016koj}. To distinguish this
approach from the standard Wilsonian RG analysis as applied
in e.g.~Refs.~\cite{Birse:1998dk,Harada:2010ba}, we will refer to it as
the large-cutoff RG-invariant (lcRG-invariant) method throughout this paper. The 
Wilsonian RG analysis addresses the running of the bare potential with
the cutoff in \emph{the infrared region}, aiming to identify 
a universal scaling behavior of perturbations around fixed points of
the RG equation. In contrast, the lcRG-invariant method of
Refs.~\cite{Nogga:2005hy,Hammer:2019poc,Valderrama:2016koj} attempts
to infer the implications of the required cutoff insensitivity of the scattering amplitude
\emph{in the deep UV region} (i.e.~for cutoff values much larger than the hard scales in the
problem) for the EFT power counting.  
 
In this paper we compare the standard approach to renormalization
as it is understood in QFT (implemented by counterterms or, equivalently, by subtracting the
loop integrals),  the Wilsonian RG method and the lcRG-invariant approach for resonant
P-wave systems  in the framework of halo EFT \cite{Bertulani:2002sz}.
The theoretical description of such systems exhibits many of the features
related to renormalization and power counting that have been under debate during the past two decades.    
In particular, it also suffers from the issue discovered in Ref.~\cite{Beane:1997pk}, where
the S-wave potential with two contact interaction terms has been iterated in the
Lippmann-Schwinger (LS) equation for the NN scattering amplitude. The
authors of that paper came to the conclusion, that the
cutoff cannot be taken beyond the hard scale of the problem
unless the effective range is non-positive.  
The solution to this problem from the EFT point of view suggested in
Ref.~\cite{Gegelia:1998xr} and reiterated in a new context 
in Ref.~\cite{Epelbaum:2018zli} is often dismissed
as irrelevant by practitioners of the lcRG-invariant approach, since  the
effective range in both the $^1$S$_0$ and $^3$S$_1$ channels may be
regarded as of natural size, so that no iterations of the subleading contact interaction
are necessary \cite{Hammer:2019poc}, see, however, Ref.~\cite{Epelbaum:2015sha}. 
This argument does not hold for resonant P-wave systems we are interested in here. 

The purpose of this paper is twofold. First, it is to be viewed as a
follow-up to a series of pedagogical papers
\cite{Epelbaum:2009sd,Epelbaum:2017byx,Epelbaum:2017tzp,Epelbaum:2018zli,Epelbaum:2019msl,Epelbaum:2020maf},
where various conceptual issues in connection with the non-perturbative
renormalization of the LS equation in the EFT context are discussed
on the example of S-waves. Secondly, we revisit the formalism and some of the conclusions of
Refs.~\cite{Bertulani:2002sz,Bedaque:2003wa,Habashi:2020qgw,Habashi:2020ofb},
where halo EFT is applied to fine-tuned S- and P-wave systems. 

Our paper is organized as follows: in Section~\ref{Halo_Nuclei_Literature}, we briefly review the 
formulations of halo EFT for S- and P-wave systems proposed in the literature and summarize the
findings to be  critically examined in the following sections.
Next, in Section~\ref{Halo_Nuclei}, we present  the formulation of
the EFT for P-wave halo systems without auxiliary dimer fields using a subtractive
renormalization scheme, while Section~\ref{WRG} addresses implications from the
Wilsonian RG analysis.  The results of our study are summarized in  Section~\ref{sec:summ}.

\section{Halo EFT with a dimer field vs.~the lcRG-invariant approach}
\label{Halo_Nuclei_Literature}

\begin{figure}[t]
\includegraphics[width=0.8\textwidth]{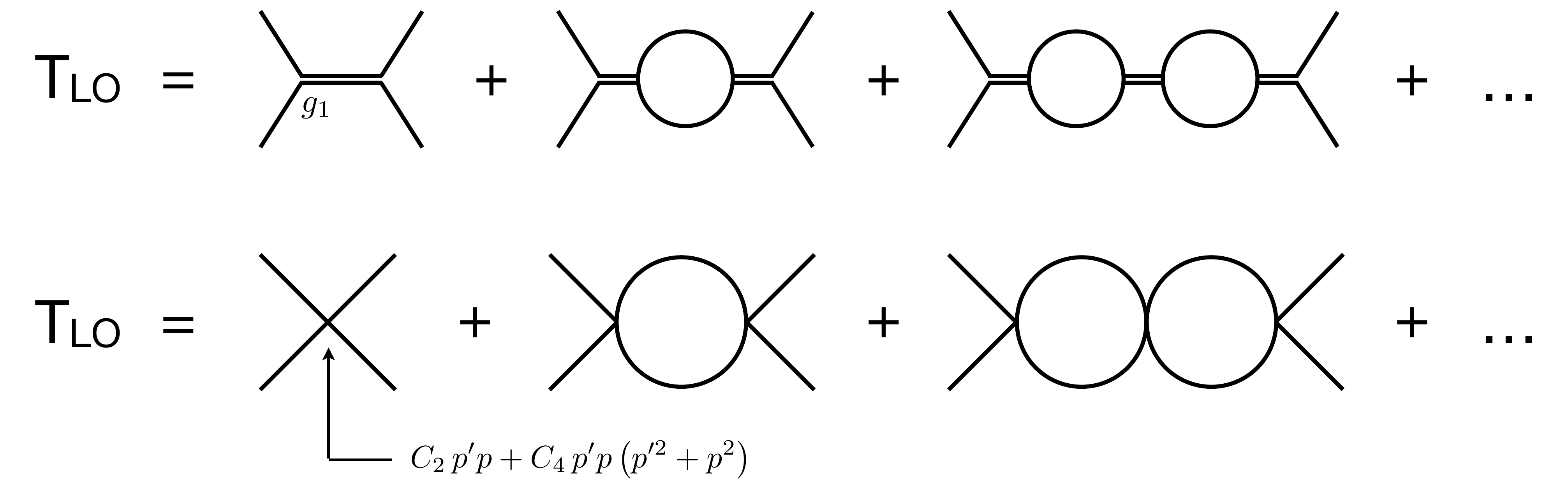}
\caption{The lowest-order amplitude for fine-tuned P-wave systems
  described in Eqs.~(\ref{pc1}), (\ref{pars}) in
the EFT with (upper panel) and without (lower panel) a dimer field. }
\label{fig:diagrams} 
\end{figure}

Consider two non-relativistic particles with interactions,  
whose (finite) range $R$ is determined by some mass scale $M_{\rm  hi} \sim 1/R$.
\footnote{Here and in what follows, we use natural
  units with $\hbar = c = 1$ unless specified otherwise.}
Near threshold, the on-shell scattering amplitude in the partial wave
with orbital angular momentum $l$ can be parameterized
in terms of the effective range expansion (ERE) \cite{Bethe:1949yr}
\be
\label{SAmp}
T(k) \propto \frac{1}{k \cot \delta - i k} \simeq
\frac{k^{2l}}{\left( - 1/a +  r k^2/2 + v_2 k^4 + \ldots  \right) - i k^{2 l +1}} \,,
\ee
where $k$ and $\delta$ denote the on-shell momentum and the phase
shift, respectively. Throughout this paper, we adopt the same naming
for the coefficients in the ERE as used for the $l=0$ case, i.e.~$a$,
$r$ and $v_i$ refer to the scattering length, effective range and the
shape parameters, respectively. If the effective range
function $k^{2l+1} \cot \delta$ does not feature poles in the
near-threshold region, the coefficients in the ERE starting from $r$ are expected to scale with the
corresponding powers of $M_{\rm hi}$, i.e.~$r \sim M_{\rm hi}^{2l-1}$,  $v_2 \sim M_{\rm hi}^{2l-3}$,
while the scattering length $a$ can take any value depending on the strength of the interaction.
In this paper, we
consider the EFT for P-wave scattering valid for momenta $k \sim M_{\rm lo} \ll M_{\rm hi}$.
We are particularly interested in fine-tuned systems,
for which the scattering amplitude in Eq.~(\ref{SAmp}) 
features poles located within the validity range of the EFT. Assuming that
the first two terms in the ERE are fine-tuned according to
\begin{equation}
1/a \sim M_{\rm lo}^3\,, \quad \quad    r  \sim M_{\rm lo}\,,  \quad \quad v_{n} \sim  M_{\rm hi}^{3-2n}\,,
\label{pc1}
\end{equation}
it follows that
the two lowest-order contact interactions in the effective two-particle potential
\begin{equation} 
V=C_2 \, p' p +C_4 \, p' p \left(p'^2+p^2\right) + \ldots \,,  
\label{potPwave}
\end{equation}
where $p \equiv | \vec p \,|$ and $p' \equiv | \vec p \,'|$ refer to
the initial and final momenta of the particles in the center-of-mass system,
need to be iterated in the LS equation to all orders \cite{Bertulani:2002sz}, see the lower
line in Fig.~\ref{fig:diagrams}.
An alternative, less fine-tuned scenario with
\begin{equation}
1/a \sim M_{\rm lo}^2 M_{\rm hi}\,, \quad \quad r \sim M_{\rm hi}\,, \quad \quad v_{n} \sim  M_{\rm hi}^{3-2n}\,,
\label{pars}
\end{equation}
has been considered in Ref.~\cite{Bedaque:2003wa}. 
The authors of both references employed the formulation of the EFT with
an auxiliary spin-$1$ dimer field following the approach developed
originally in Ref.~\cite{Kaplan:1996nv} for the case of NN  S-wave
scattering, see the upper panel in Fig.~\ref{fig:diagrams}.
For applications of  EFTs with auxiliary fields
to  nuclear systems see
e.g.~Refs.~\cite{Gelman:2009be,Alhakami:2017ntb,Schmidt:2018vvl, Ji:2014wta,Ryberg:2017tpv,Soto:2007pg}.  
Notice that the UV divergences in the dimeron self-energy at leading order are cancelled by
the counterterms generated by the bare particle-dimeron coupling
constant $g_1$ and the residual dimeron mass $\Delta_1$, see
Refs.~\cite{Bertulani:2002sz,Bedaque:2003wa} for details.

Recently, also highly fine tuned S-wave systems with shallow
resonances have been analyzed in the EFTs 
without \cite{Habashi:2020qgw}  and with \cite{Habashi:2020ofb} 
an auxiliary dimer field assuming the scaling behavior 
$a \sim r \sim 1/M_{\rm lo}$, $v_{n} \sim  M_{\rm hi}^{1-2n}$, so that the first two terms in
the ERE are of the same order as the unitarity term $- i k$.
The required deviation from NDA for the first two
terms in the ERE represents a minimal condition needed to generate low-lying
resonance states in S-waves. For the formulation without 
auxiliary fields, the authors of Ref.~\cite{Habashi:2020qgw}
considered energy-independent contact interactions using the lcRG-invariant
approach. That is, the expression for the on-shell amplitude resulting
from the iteration of the potential $C_0 + C_2 (p'^2 + p^2)$ in the
cutoff-regularized LS equation is matched to the first two terms in
the ERE for arbitrarily large values of the UV cutoff
  $\Lambda$. In fact, exactly the same approach was used a
long time ago by Beane et al.~\cite{Beane:1997pk} to describe NN scattering. 
As already pointed out in the introduction, taking $\Lambda \gg M_{\rm hi}$ leads to
complex values for the bare LECs $C_0 (\Lambda )$, $C_2 (\Lambda )$ unless the effective
range is negative. This observation is a manifestation of the well known Wigner bound
\cite{Wigner:1955zz}, a constraint on the effective range placed by the range of the
interaction $R$, $r \leq 2 R [1 + \mathcal{O} (R/a)]$, that relies on
causality and unitarity.  For a generalization of the Wigner bound to
higher partial waves and arbitrary dimensions, see
Ref.~\cite{Hammer:2010fw}. So, how can then the positive experimental
values for the effective range in the neutron-proton $^1$S$_0$ and
$^3$S$_1$ channels, namely $r=2.75(5)$~fm and $r=1.759(5)$~fm \cite{Dumbrajs:1983jd},
be reconciled with the EFT? As pointed out in Refs.~\cite{Gegelia:1998xr,Epelbaum:2018zli} and will
be demonstrated in the next section for the case of P-wave
scattering, the constraint on the value of $r$ in the lcRG-invariant
formulation of the EFT is an artifact of the amplitude being only
partially renormalized prior to taking the $\Lambda \to \infty$ limit.
The issue with the Wigner bound becomes
irrelevant once the amplitude is properly renormalized using e.g.~a
subtractive scheme regardless of whether the $C_2$-term is treated in
perturbation theory or non-perturbatively. It also does not pose a
problem in both pionless and chiral EFTs for NN scattering
if the UV cutoff is kept of the order of the corresponding hard scale
as done e.g.~in Refs.~\cite{Epelbaum:2014sza,Entem:2017gor,Reinert:2017usi}.
Furthermore, if one assumes $r \sim 1/M_\pi$ for both S-wave NN channels, the
range corrections can be taken into account perturbatively in pionless
EFT with no restrictions on the value of $r$, regardless of the
employed cutoff value.  

On the other hand,  the issue with the Wigner bound cannot be avoided in the 
lcRG-invariant EFT for shallow S-wave resonances. The
authors of Ref.~\cite{Habashi:2020qgw} therefore conclude that
\emph{``renormalization at leading order forces the effective range to be
negative''}. Since a negative effective range admits only at most one
solution of the equation $-1/a + r k^2 /2 + i k =0$ in the upper half
of the complex momentum plane, no unphysical poles in the amplitude
corresponding to ${\rm Re} \, k \neq 0$, ${\rm Im} \, k > 0$ can
appear. The authors thus come to the conclusion that \emph{``renormalization
automatically incorporates the causality constraint that a resonance
represents decaying, not growing, states''}  \cite{Habashi:2020qgw}.

Following the approach of Ref.~\cite{Habashi:2020qgw}, we now 
apply the lcRG-invariant EFT formulation without dimer fields
to the case of resonant P-wave scattering. To this aim, we solve the LS equation
for the off-shell P-wave  amplitude 
\be
T(p',p,k)=V(p',p)+  m \int_0^\Lambda \frac{l^2 d l}{2\pi^2} \, \frac{V(p,l) \;T(l,p',k)}{k^2-l^2+i\,\epsilon}\,,
\label{eqLS}
\ee
where the bare potential is given in Eq.~(\ref{potPwave}), and we have
introduced a sharp cutoff to render the appearing integrals UV-convergent,
\be
\label{IntJ}
J_n (k) = \int_0^\Lambda  \frac{l^2 dl}{2 \pi^2} \, \frac{m \, l^{n-1}}{k^2 - l^2 +
i \epsilon} = I_n + k^2 I_{n-2} + \ldots + k^{n-3} I_3 + k^{n-1} I(k)  , \quad \quad n = 3, 5, \ldots\,, 
\ee
where the superscript $n$ denotes the degree of divergence and the integrals $I_n$ and $I(k)$ are defined
via 
\bea
\label{IntI}
I_n &=& -m \int_0^\Lambda \frac{l^2 dl}{2 \pi^2} \, l^{n-3} = - \frac{m
  \Lambda^n}{2n  \pi^2} , \quad \quad n = 1, 3, 5, \ldots\,, \nonumber \\[3pt]
I (k) &=& \int_0^\Lambda  \frac{l^2 dl}{2 \pi^2} \, \frac{m}{k^2 - l^2 +
i \epsilon} = I_1 - i \frac{mk}{4 \pi} - \frac{m k}{4 \pi^2 }
\ln \frac{\Lambda - k}{\Lambda + k}\,.
\eea
We then obtain for the on-shell amplitude $ T(k) \equiv T(k, k, k)$:
\be
\label{BareScattAmpl}
\frac{k^2}{T(k)} =-I(k) \,  k^2-I_3 + 
\frac{\left(C_4 I_5-1\right){}^2}{C_4 \left(k^2 \left(2-C_4 I_5\right)+C_4
   I_7\right)+C_2}\,.
\ee
Notice that for the sake of compactness, we have suppressed the dependence of
the integrals $I(k)$, $I_n$ and the bare LECs $C_2$, $C_4$ on the
cutoff $\Lambda$. To perform (implicit) renormalization, we express
the bare LECs $C_2(\Lambda )$, $C_4 (\Lambda )$  in terms of the
scattering length and effective range by expanding
Eq.~(\ref{BareScattAmpl}) in powers of $k^2$ and matching the first
two coefficients to the inverse of Eq.~(\ref{SAmp}). Following the
lcRG-invariant scheme, we take the $\Lambda \to \infty$ limit to arrive
at the cutoff-independent expression for the scattering amplitude
\begin{equation}
T(k) = - \frac{4\pi}{m} \frac{k^2}{ - \frac{1}{a} +\frac{r}{2} k^2 -i k^3}\,.
\label{amplPW}
\end{equation}
While this result looks satisfactory, the expressions for the bare
LECs $C_2$,  $C_4$ in terms of the scattering
length and effective range  have the form\footnote{The matching
  equations also admit another solution with the ``$-$'' sign in
  front of the square roots. This solution is, however, incompatible
  with the loop expansion of the amplitude and, therefore, has to be discarded as unphysical.}
\bea
\label{ccs}
\frac{m}{10 \pi^2} C_2 (\Lambda) &=&\frac{64 a^2 \Lambda ^6 + 10
    \sqrt{5 \alpha } \left(3 \pi -2 a
   \Lambda ^3\right)+3 \pi  a \Lambda ^3 \left(15 a \Lambda ^2 r+158\right)-450 \pi
   ^2}{7 \Lambda ^3  \left(16 a^2 \Lambda ^6+3 \pi  a \Lambda ^3 \left(3 a
     \Lambda ^2 r+20\right)-45 \pi ^2\right)}\,, \nonumber \\
\frac{m}{10 \pi^2} C_4 (\Lambda) &=& \sqrt{\frac{5}{\alpha}}
  \frac{\left(3 \pi -2 a \Lambda ^3\right)}{\Lambda^5}- \frac{1}{\Lambda^5}\,,
\eea
where we have introduced
$\alpha = -16 a^2 \Lambda ^6-3 \pi  a \Lambda ^3 \left(3 a \Lambda ^2 r+20\right)+45 \pi ^2$.
Thus, both bare coupling constants  become
complex  for sufficiently large values of the cutoff $\Lambda$. This
observation is in line with the causality bound $r \leq -2/R \,  ( 1 + \mathcal{O} (R^3/a))$
obtained in Ref.~\cite{Hammer:2010fw} if 
the range of the interaction $R$ is identified with $1/ \Lambda$. Taking
the renormalizability requirement of the lcRG-invariant approach seriously as done, e.g., in
Refs.~\cite{Hammer:2019poc,Habashi:2020qgw}, one is forced to conclude that resonant
P-wave systems specified by Eqs.~(\ref{pc1}), (\ref{pars}) cannot be described in an EFT without an auxiliary
dimer field. As we will show in the next section, the problem actually lies
in the procedure of the lcRG-invariant approach rather than in the EFT itself. 

As already mentioned in the introduction, resonant P-wave
systems have also been examined in the EFT with auxiliary dimer fields
\cite{Bertulani:2002sz,Bedaque:2003wa,Gelman:2009be,Alhakami:2017ntb},
see Ref.~\cite{Hammer:2017tjm} for a review article.
The EFT formulation employed in these studies may, however, admit unphysical solutions. For example, one may
encounter shallow poles in the upper half plane \cite{Habashi:2020qgw}.
The possible appearance of unphysical solutions makes the mismatch between the lcRG-invariant and
the dimer-field EFT formulations less evident. It is, however, easy to construct
a simple example that leads to shallow P-wave states in agreement
with the assignment of Eq.~(\ref{pars}) and is
compatible with causality and unitarity. For this, we simply take the
solution for the amplitude obtained in the  lcRG-invariant
approach but keep the cutoff $\Lambda$ finite of the order of $\Lambda \sim M_{\rm hi}$,
as advocated in
Refs.~\cite{Epelbaum:2009sd,Epelbaum:2017byx,Epelbaum:2018zli,Epelbaum:2019kcf,Lepage:1997cs}.    
Substituting the values of $C_2( \Lambda )$ and  $C_4( \Lambda )$ from
Eq.~(\ref{ccs}) into Eq.~(\ref{BareScattAmpl}), the effective range function is found
to be 
\be
k^3 \cot \delta = - \frac{1}{a} + \frac{1}{2} r k^2 -
\frac{k^4}{2 \pi }
\left(\frac{3   (4 \Lambda +\pi  r)^2}{
  6 \pi a^{-1} - 4  \Lambda^3 + 3  k^2 (4 \Lambda + \pi r)}+\frac{2}{k} \ln \frac{\Lambda
     -  k}{\Lambda + k
   }\right)\,.
\ee
For $\Lambda \sim M_{\rm hi}$, the cutoff dependent
coefficients in  the ERE  terms $\sim k^{2n}$, $n = 2, 3, \ldots$, are beyond the accuracy
of the LO approximation for the assumed power counting scenario. The condition $C_{2,4} (\Lambda)
\in \mathbb{R}$ translates into the following restriction on the effective range 
\be
r \le \frac{5 \pi}{a^2 \Lambda^5} - \frac{20}{3 a \Lambda^2} -
\frac{16 \Lambda}{9 \pi} \lesssim -
\frac{16 \Lambda}{9 \pi} \,,
\ee
where the second inequality is valid for the assumed enhanced values of
the scattering length. Thus, the considered example is compatible with
the scenario suggested in Eq.~(\ref{pars}) and describes a P-wave
system that exhibits a deeply bound state outside of the EFT validity
range and either a shallow narrow resonance for $a < 0$ or a
combination of shallow bound and virtual states for $a > 0$, see 
Refs.~\cite{Bertulani:2002sz,Ji:2014wta,Hammer:2017tjm} for a related
discussion. This situation cannot
be accommodated by the lcRG-invariant approach.

\section{Subtractively renormalized halo EFT for P-wave scattering}
\label{Halo_Nuclei}

We now renormalize the amplitude  in
Eq.~(\ref{BareScattAmpl}) following the standard procedure in QFT (and
EFT) by subtracting \emph{all} UV divergences prior to removing the
regulator by taking the limit $\Lambda \to \infty$. For the case
at hand, this corresponds to taking into account contributions of an
infinite number of counterterms, see
Refs.~\cite{Gegelia:1998xr,Gegelia:1998gn,Epelbaum:2017byx,Epelbaum:2020maf,Ren:2020wid,Ren:2021yxc}
for a related discussion. Specifically, we first separate out power-like UV divergences
in the appearing integrals in the most  general way via 
\bea
\label{IntIRen}
I_n &=& - m
\int_{0}^{\mu_n} \frac{l^2 dl}{2 \pi^2} l^{n-3} - m \int_{\mu_n}^\Lambda \frac{l^2 dl}{2 \pi^2} l^{n-3} 
\equiv  I_n^R (\mu_n)
+ \Delta_n (\mu_n )
\,, \quad \quad
\mbox{with} \quad n = 1, 3, 5, \ldots\,,\nonumber \\
I (k) &\equiv&  I^R (k, \mu_1 ) - \Delta_1 (\mu _1)\,,
\eea
where $\mu_n$ denotes the corresponding subtraction scales. We then
renormalize the scattering amplitude in
Eq.~(\ref{BareScattAmpl})  by simultaneously replacing the integrals $I_n$ and $I (k)$
with  $I_n^R (\mu_n)$ and $I^R (k, \mu_1)$ and the bare coupling
constants $C_2$ and $C_4$ with the corresponding
$\mu_n$-dependent renormalized couplings   $C_2^R$ and $C_4^R$,
respectively. As will be shown below, by doing so we implicitly take into account the
contributions of an infinite number of counterterms.
Since the renormalized amplitude depends only on
UV-convergent integrals, we can now safely take the limit $\Lambda \to \infty$.
Fixing the renormalized LECs by the requirement to
reproduce the scattering length and effective range leads to
our final result for the subtractively renormalized effective range function expressed in
terms of physical parameters:
\be
\label{ERFRen}
k^3 \cot \delta = - \frac{1}{a} + \frac{1}{2} r k^2 -\frac{k^4}{2 \pi}
\frac{
3 \left(4 \mu _1+\pi  r\right){}^2}{6 \pi a^{-1} - 4  \mu_3^3 + 3  k^2 (4
 \mu_1 + \pi r)}\,.
\ee

It is instructive to discuss some of the qualitative features of the
obtained result. We first observe  that the renormalized scattering
amplitude depends on the subtraction scales $\mu_1$ and $\mu_3$. This can
be traced back to non-renormalizability of the potential in
Eq.~(\ref{potPwave}), which reflects the fact that not all UV divergences
generated by the loop expansion of the amplitude are cancelled by 
counterterms stemming from $C_2$ and $C_4$. As already mentioned
above, our renormalization procedure is, in fact, equivalent to taking
into account an infinite number of scale-dependent counterterms, while 
utilizing a specific (fixed) choice for the corresponding scale-dependent
renormalized LECs. To elaborate on this point, consider an EFT
formulation that allows for energy-dependent contact interactions.
In such a case, one can easily obtain the
expression for the bare potential corresponding to the renormalized
one $C_2^R p' p + C_4^R p' p (p'^2 + p^2)$, which incorporates {\it all}
counterterms, in a closed
form:\footnote{To do so, one can start from a separable potential of the form
$V = p' p f_1 (k) + p p' (p'^2 + p^2) f_2 (k) + p'^3p^3 f_3 (k)$ and determine
the functions $f_i (k)$  from matching the
off-shell $T$-matrix, obtained by solving the  
cutoff-regularized LS equation, to the subtractively renormalized
off-shell $T$-matrix.} 
\bea
\label{BarePot}
V&=&C_2^R p p' + C_4^R p p' \left(p^2+p'^2\right) + \text{counterterms} \nonumber \\[4pt]
&=&p p' \frac{C_2^R + C_4^R \left(p^2+p'^2\right)-
  \hbar (C_4^R)^2 \, p^2 p'^2 \left(J_3-J_{3}^R\right)-\hbar (C_4^R)^2
  \left[J_7 -J_{7
     }^R  -\left(J_5-J_{5}^R\right)
    \left(p^2+p'^2\right)\right]}{1+ \hbar C_2^R  \left(J_{3}-J_3^R\right)
+2 \hbar C_4^R  \left(J_{5}-J_5^R\right)
+  \hbar^2 (C_4^R)^2 \big[\left(J_5-J_{5}^R\right)^2-\left(J_3-J_{3}^R\right) \left(J_7-J_{7
  }^R\right)\big]} \nonumber \\
&=&  p p' \frac{N}{D}\,,
\eea
where 
\bea
\label{ND}
N&=&42 C_2^R +
C_4^R \bigg(42
   \left(p^2+p'^2\right) + \hbar \tilde C_4^R \bigg\{70 \left(k^2-p^2\right) \left(k^2-p'^2\right)
   \left[3 k^2 (\Lambda - \mu_1)  + \Lambda ^3-2 \mu _3^3\right] +30
   \left(\Lambda ^7-\mu _7^7\right)
   \nonumber \\
   &+& 42 \left(\Lambda ^5-\mu
     _5^5\right) \left(k^2-p^2-p'^2\right) \bigg\}\bigg)\,, \nonumber
   \\
D&=&  42 - 14 \hbar \tilde C_4^R \left[5 k^2 \left(\Lambda
    ^3-\mu _3^3\right)+15 k^4 \left(\Lambda -\mu _1\right)+3 \left(\Lambda ^5-\mu
    _5^5\right)\right] \left[2 - \hbar \tilde C_4^R
  \left(\Lambda ^5- \mu _5^5\right)\right]\nonumber \\
& -&10 \left[3 k^2 \left(\Lambda - \mu _1\right)+\Lambda ^3-\mu
  _3^3\right] \left[7 \hbar \tilde C_2^R + 5 (\hbar \tilde C_4^R)^2 
   \left(\Lambda ^7-\mu _7^7\right)\right]\,.
\eea
In the above expressions, $J_n \equiv  J_n (k)$ are the
cutoff-regularized integrals defined in Eqs.~(\ref{IntJ}), (\ref{IntI}), while
$J_n^R \equiv  J_n^R (k, \mu_i)$ refer to the corresponding
cutoff-dependent but UV-convergent renormalized integrals, obtained by
replacing $I_n$ in Eq.~(\ref{IntJ}) with $I_n^R (\mu_n)$ defined in
Eq.~(\ref{IntIRen}).  Furthermore, we have introduced 
$\tilde C_n^R   \equiv  m C_n^R/(10 \pi^2)$ to simplify the notation and 
retained the factors of $\hbar$ to facilitate the interpretation of
our results in terms of the loop expansion.
The dependence of the bare potential in Eq.~(\ref{BarePot})
on the combinations of the integrals $J_n - J_n^R$ only is
consistent with the employed renormalization procedure.  
Solving the cutoff-regularized
LS equation with the potential in Eq.~(\ref{BarePot}), one can 
verify that the resulting scattering amplitude $T (p', p, k )$
matches exactly the subtractively
renormalized one, with $\lim_{\Lambda \to \infty} {\rm Re} \{-4 \pi k^2/[ m T(k)] \}$ coinciding with 
Eq.~(\ref{ERFRen}). 

After these preparations, it is easy to explicitly verify the
equivalence of the employed renormalization procedure and the 
standard QFT/EFT renormalization technique based on splitting the bare
coupling constants into the renormalized ones and counterterms. 
That is, we start with the potential written in terms of bare LECs,
which involves an infinite number of  contact interactions
(some of which are redundant)
\be
\label{PotFull}
V = p'p \left( C_2 + C_2^2 k^2 + C_2^4 k^4 + \ldots \right)
+ p'p (p'^2 + p^2) \left( C_4 + C_4^2 k^2 + C_4^4 k^4 + \ldots \right)
+ p'^3p^3 \left( C_6 + C_6^2 k^2 + C_6^4 k^4 + \ldots \right) + \ldots\,,
\ee
where the ellipses refer to terms with higher powers of $p$, $p'$ and
$k$. The subscripts (superscripts) of the LECs accompanying various  terms
denote the powers of the off-shell momenta $p'$, $p$ (on-shell momentum $k$). For resonant P-wave
systems described by Eqs.~(\ref{pc1}), (\ref{pars}),
the LO scattering amplitude is obtained by resumming 
the $C_2^R$- and the $C_4^R$-contributions as explained below\footnote{Alternatively and
  equivalently, one can use the $C_2^2$-vertex instead of the
  $C_4$-one or their linear combination.},
while insertions
of the interactions with higher powers of momenta or energy are 
suppressed by powers of $M_{\rm lo}/M_{\rm  hi}$
for an appropriate choice of renormalization conditions to be
specified below. Therefore, we set the renormalized LECs accompanying
higher-order terms in the potential to zero as appropriate at LO:
\be
\label{RenChoice}
C_l^R (\mu_i)
= C_n^{m,  R} (\mu_i) = 0 \quad  \text{for} \;\;
\forall n,m \;\;
\text{and} \;\;
l \ge 6 
\,.
\ee
The scattering amplitude can be calculated from iterations of the
cutoff-regularized LS equation with the potential in
Eq.~(\ref{PotFull}) at any loop order. Renormalization is accomplished in the
usual way by splitting the unobservable bare
LECs into the renormalized ones and (scheme-dependent) counterterms,
which depend on the renormalized LECs. For all renormalized LECs
being set to zero except for $C_2^R$ and $C_4^R$, this splitting has
the form  
\bea
\label{ctSplit}
C_2 (\Lambda ) &=& C_2^R (\mu_i ) + \sum_{L=1}^\infty \hbar^L
\Delta_{C_2}^{(L)} (\Lambda, \mu_i) \,, \nonumber
\\
C_4 (\Lambda ) &=& C_4^R (\mu_i ) + \sum_{L=1}^\infty \hbar^L
\Delta_{C_4}^{(L)} (\Lambda, \mu_i) \,, \nonumber
\\
C_6 (\Lambda ) &=&  \sum_{L=1}^\infty \hbar^L
\Delta_{C_6}^{(L)} (\Lambda, \mu_i) \,, \nonumber
\\
C_n^m (\Lambda ) &=&  \sum_{L=1}^\infty \hbar^L
\Delta_{C_n^m}^{(L)} (\Lambda, \mu_i) \,, 
\eea
while $C_{\ge 8} (\Lambda ) = C_{\ge 8}^m (\Lambda ) =0$.  The
explicit expressions for the counterterms on the right-hand sides of the above equations
can be read off from Eqs.~(\ref{BarePot}), (\ref{ND}). For example, for
the counterterms in the first line of Eq.~(\ref{ctSplit}), one has 
\bea
&&\sum_{L=1}^\infty \hbar^L
\Delta_{C_2}^{(L)} \\
&&=  \hbar \frac{  21 C_2^R \tilde C_4^R
  \left(\Lambda ^5 - \mu _5^5\right) \left[2 - \hbar \tilde C_4^R 
    \left(\Lambda ^5 - \mu _5^5\right)\right]
+ 15 C_4^R \tilde C_4^R \left(\Lambda ^7-\mu _7^7\right)
 + 5 C_2^R \left(\Lambda ^3- \mu _3^3\right)
   \left[5 \hbar (\tilde C_4^R)^2 \left(\Lambda ^7-\mu _7^7\right)+7
     \tilde C_2^R\right] }{21-21 \hbar \tilde C_4^R  \left(\Lambda
     ^5-\mu _5^5\right) \left[2 - \hbar \tilde C_4^R  \left(\Lambda ^5
       - \mu
       _5^5\right)\right]-5 \hbar \left(\Lambda ^3 - \mu _3^3\right)
   \left[5 \hbar  (\tilde C_4^R)^2 \left(\Lambda ^7-\mu _7^7\right)+7 \tilde C_2^R\right]}\,,
\nonumber
 \eea
where we have suppressed the subtraction-scale dependence of the
renormalized LECs $C_2^R$ and $C_4^R$. Being expressed in terms of the
renormalized LECs as described above,  the
scattering amplitude at any loop order $\hbar^L$ involves only
UV-convergent integrals, so that one can safely take
the limit $\Lambda \to \infty$. The LO amplitude is obtained by
resumming the finite contributions to all loop orders. Setting $p' = p = k$
and fixing $C_2^R$ and $C_4^R$ from matching the first two terms
in the ERE leads to the expression given in Eq.~(\ref{ERFRen}). The above
considerations make it clear that the residual dependence of the amplitude
on the scales $\mu_1$, $\mu_3$ is induced by our choice
for the renormalized coupling constants of higher-order contact
interactions in Eq.~(\ref{RenChoice}). Notice further that
contrary to what is claimed in Refs.~\cite{Bertulani:2002sz,Habashi:2020qgw,Habashi:2020ofb},
renormalization by itself imposes no constraints on the values of the
coefficients in the ERE. 

So far, we have left open the question of the choice of the
subtraction scales $\mu_i$, which plays a key role in setting up a
self-consistent power counting.
For fine-tuned S-wave systems near the
unitary limit with $a \sim 1/M_{\rm lo}$, it is
possible to choose all subtraction scales of the order of the soft scale,
i.e.~$\mu_i \sim M_{\rm lo}$. This  leads to manifest
power counting for loop diagrams, commonly referred to as the KSW
scheme \cite{Kaplan:1998tg}. For this choice of the renormalization
conditions, the LEC accompanying the LO (i.e., derivative-less) contact interaction
is enhanced compared to NDA, $C_{0}^R  \sim M_{\rm lo}^{-1}$, and the LO amplitude
$\sim M_{\rm lo}^{-1}$ is  generated by resumming all possible bubble diagrams constructed
from the $C_{0}^R$-vertices, which all scale as  $\sim M_{\rm lo}^{-1}$.
Higher-order corrections to the amplitude are enhanced by
$\sim M_{\rm lo}^{-2}$ relative to NDA and can be taken into account
perturbatively. They stem from dressed higher-order contact interactions 
accompanied with enhanced LECs. Notice that while the renormalization conditions $\mu_i
\sim M_{\rm lo}$ seem to permit choosing $\mu_i = 0$, which would be the case if one would use dimensional
regularization (DR) in combination with the minimal subtraction (MS) or modified minimal
subtraction scheme ($\overline{\mbox{MS}}$), setting $\mu_1 = 0$
results in the EFT expansion that has
zero radius of convergence for $a \to \infty$ \cite{Kaplan:1996xu,Beane:1997pk}. The issue can be
avoided using a subtractive renormalization scheme \cite{Gegelia:1998xr} or DR in
combination with the power divergence subtraction (PDS) scheme to explicitly 
account for linear divergences by subtracting poles in $d = 3$
space-time dimensions \cite{Kaplan:1998tg}.  Alternatively
to the KSW approach, a self-consistent power counting scheme for
S-wave systems with a large scattering length is obtained by
setting $\mu_1  \sim M_{\rm hi}$ while  keeping $\mu_3 \sim \mu_5 \sim \ldots \sim M_{\rm lo}$
\cite{Epelbaum:2017byx}. This choice of the renormalization conditions
leads to Weinberg's power counting with 
all LECs scaling according to NDA. All bubble diagrams constructed
from the LO contact interactions scale individually as $\mathcal{O}(1)$, but their
resummed contribution is enhanced by $M_{\rm lo}^{-1}$ as a result of fine-tuning the
LEC $C_0^R$. Higher-order corrections are again generated perturbatively from dressed contact
interactions with increasing number of derivatives.  

For the case of resonant P-wave scattering we are interested in here, one
may expect the choice of renormalization conditions to be even
more delicate due to the even stronger amount of fine tuning.
Indeed, a closer look
at Eq.~(\ref{ERFRen}) reveals that one \emph{must} choose $\mu_3 \sim M_{\rm  hi}$ since setting
$\mu_3 \sim M_{\rm lo}$ would lead to poles in
the effective range function\footnote{Such poles correspond to the
  phase shift crossing zero and do not contradict any fundamental
  principle. They do, however, restrict the range of convergence of the
  ERE.} located at $k \sim M_{\rm lo}$, thereby resulting in
enhanced values of the coefficients in the ERE in contradiction with
the assumed scenarios in Eqs.~(\ref{pc1}), (\ref{pars}). Consequently, no
KSW-like scheme is possible for resonant P-wave systems under
consideration.\footnote{In fact, the same issue appears in fine-tuned
  S-wave systems as well, if the effective range term is treated  non-perturbatively \cite{Epelbaum:2015sha}.  }
A self-consistent Weinberg-like scheme with manifest power counting
for renormalized loop diagrams and all LECs $C_n^R$, $C_n^{m,  R}$ scaling according to
NDA emerges if we set  $\mu_{5}  \sim  \mu_{7} \sim  \mu_{9} \sim \ldots \sim  M_{\rm lo}$.
The remaining scale $\mu_1$ can be chosen either as $\mu_1 \sim M_{\rm hi}$ or  $\mu_1 \sim M_{\rm lo}$
as will be discussed below.

\begin{figure}[t]
\includegraphics[width=\textwidth]{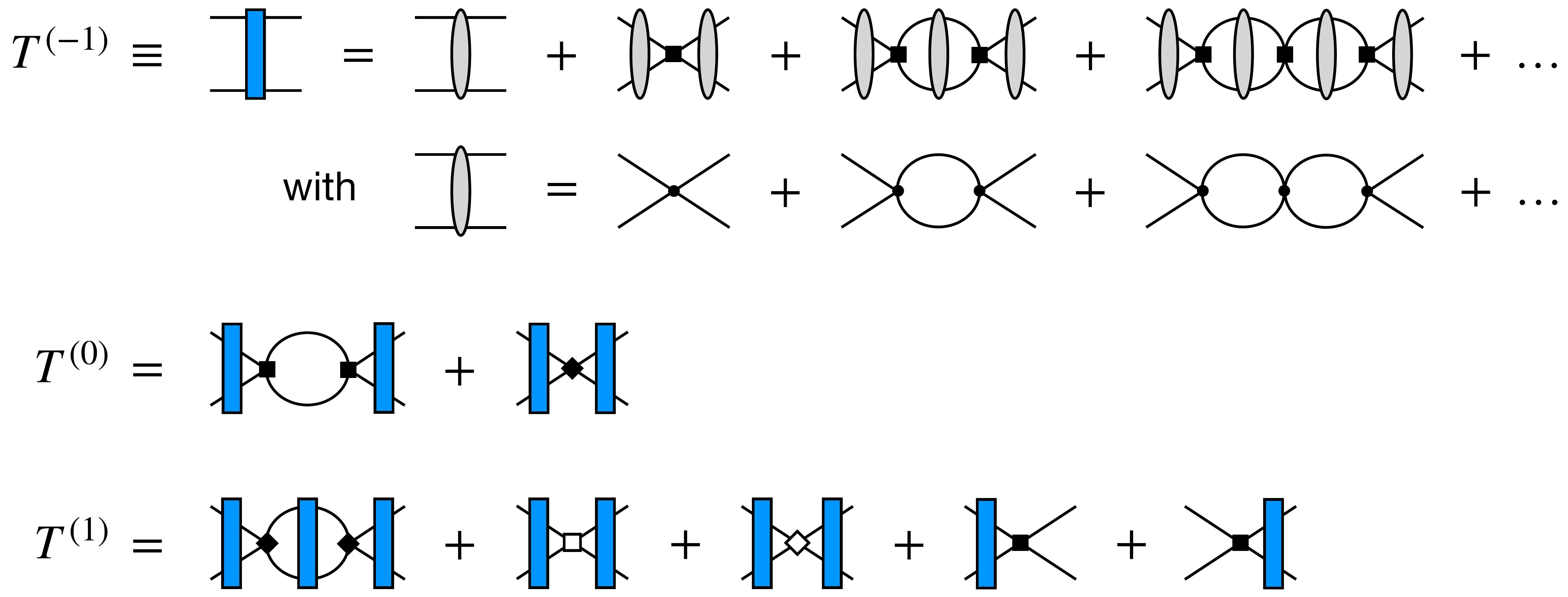}
\caption{The leading, subleading and sub-subleading contributions to
  the P-wave scattering amplitude for the scenario of
  Eq.~(\ref{pc1}). Solid dots, rectangles and diamonds refer to the
  $C_2$-, $C_4$- and $C_6$-vertices, respectively. Open rectangles and
  open diamonds denote corrections to the $C_4$- and $C_6$-vertices
proportional to  $\delta C_4^{(3)}$ and $\delta C_6^{(1)}$ as
explained in the text. For all diagrams, the two-particle Green's
functions refer to the usual nonrelativistic free resolvent operator
as appears in Eq.~(\ref{eqLS}). }
\label{fig:diagrams_pc1} 
\end{figure}

It is instructive to see how the P-wave  scattering amplitude is obtained
in terms of
diagrams for both scenarios specified in
Eqs.~(\ref{pc1}) and (\ref{pars}). Here and in what follows, we set
the renormalization conditions as
\be
\label{RenormCond}
\mu_1 \sim \mu_3 \sim M_{\rm hi}, \quad \quad
\mu_5 = \mu_7 = \ldots = 0\,. 
\ee
Notice that setting the scales $\mu_{\ge 5} = 0$ is not necessary and
done solely to keep the resulting expressions simple. The low-energy
expansion for the scattering amplitude for the doubly fine-tuned scenario
of Eq.~(\ref{pc1}) is visualized in Fig.~\ref{fig:diagrams_pc1}. 
For the employed renormalization conditions, a two-particle scattering diagram
made out of $V_i$ vertices of type $i$ with all LECs scaling according
to NDA starts contributing to the amplitude at order $\sim M_{\rm lo}^n$ with 
\be
n = 2 + \sum_i V_i (d_i - 2 ) \,,
\ee
where $d_i$ is the power of momenta for a vertex of type $i$. Consequently, all diagrams
constructed solely from the lowest-order vertices $\propto C_2$ and shown in the second line of
Fig.~\ref{fig:diagrams_pc1} contribute at the same order $n=2$ and, therefore,
 must be resummed. Their resummed off-shell contribution $T_{C_2}(p', p, k)$ has the form
\be
T_{C_2}
(p', p, k) = \frac{12\pi^2 C_2^R  p' p }{12 \pi^2 + m C_2^R ( 3i  \pi
  k^3 + 6 \mu_1 k^2 + 2 
  \mu_3^3 )} \,.
\ee
Since no other diagram can contribute to the scattering length 
(for the employed renormalization conditions), the exact value of
the LEC  $C_2^R$ can be determined from matching $T_{C_2}(k, k, k) $
to $\big[T (k)/k^2 \big]_{k = 0}= 4 \pi a/m$, yielding
\be
\label{C2R}
C_2^R (\mu_3) = \frac{12\pi^2 }{m (3\pi a^{-1}- 2 \mu_3^3 )}
\,.
\ee
While $C_2^R$ is of natural size, its value had to be fine-tuned to
reproduce $a^{-1} \sim M_{\rm lo}^3$, leading to the amplitude
\be
\label{TC2}
T_{C_2}
(p', p, k) = -\frac{4\pi}{m} \frac{p' p}{-\frac{1}{a} -
  \frac{2\mu_1}{\pi} k^2  - i k^3}
\; \sim \; \mathcal{O} (1)
\,,
\ee
which is enhanced by two inverse powers of the soft scale relative to the
expectation based on NDA. As a consequence, all
diagrams made out of $m$ subleading vertices $\propto C_4$ and $m+1$
insertions of $T_{C_2}$, see the first line of
Fig.~\ref{fig:diagrams_pc1}, are enhanced and appear at the same order
$\sim \mathcal{O} (1)$.  Their resummed contribution defines
the LO amplitude $T^{(-1)} (p', p, k) \sim \mathcal{O} ( M_{\rm lo}^{-1} )$, which is
additionally enhanced by one inverse power of $M_{\rm lo}$ as a result
of the fine tuning of the value of $C_4^R$,
\be
\label{C4leading}
C_4^R (\mu_1, \mu_3) = \frac{9\pi^2 (\pi r + 4 \mu_1) }{4m \mu_3^6}
\,,
\ee
needed to reproduce the effective range $r \sim M_{\rm lo}$.
Notice that similarly to  $C_2^R (\mu_3)$,  $C_4^R (\mu_1, \mu_3)  \sim \mathcal{O} (1)$ is also
consistent with NDA.  The resulting LO amplitude reads
\bea
\label{LO}
T^{(-1)} (p', p, k) &=& - \frac{4\pi}{m} \, \frac{p'p}{- \frac{1}{a} + \frac{1}{2}
r k^2 - i k^3 + \frac{3 \mu_1 (\pi r + 4 \mu_1)}{\pi \mu_3^3}k^4
+ i \frac{3  (\pi r + 4 \mu_1)}{2 \mu_3^3}k^5 + \mathcal{O} (k^6)}
\nonumber \\
&=& - \frac{4\pi}{m} \, \frac{p'p}{- \frac{1}{a} + \frac{1}{2}
r k^2 - i k^3} + \mathcal{O} ( 1 )\,.
\eea
We emphasize that the obtained LO amplitude is, by construction, RG
invariant at the considered level of accuracy since scale-dependent terms
appear at order $\mathcal{O}(1)$. 

Corrections to the LO amplitude emerge from diagrams involving
insertions of vertices with a larger number of derivatives.
As already pointed out above, the choice of higher-order operators is not unique. 
In particular,  one can use any of the order-$M_{\rm lo}^6$ operators
$p'^3 p^3$, $p' p (p'^4+ p^4)$, $p' p k^4$ or  $p'p (p'^2+ p^2) k^2$ as they are
all equivalent on-shell. In addition to
vertices involving higher powers of momenta, one has to take into
account interactions emerging from higher-order corrections to the
renormalized LECs. For example, the expression for $C_4^R$ in
Eq.~(\ref{C4leading}) is only correct up to terms of order $\sim M_{\rm lo}^2 $,
and one still needs to include the contributions
from order-$M_{\rm lo}^m$ corrections $\delta C_4^{ (m)}$ with $m = 3, 4, 5, \ldots$. Alternatively
and equivalently (up to higher-order terms), one can account for such corrections
without introducing new vertices by re-adjusting the LECs 
$C_n^{R}$ at orders beyond the one they start contributing to the
amplitude.

For the employed renormalization conditions, a diagram made out of $V_i$
vertices of type $i$ and order $d_i \ge 4$ with $N$ insertions of the LO
amplitude in Eq.~(\ref{LO}) contributes at order $M_{\rm lo}^n$ with
\be
\label{PowC1}
n = 2 - 3N + \sum_i V_i (d_i - 2 )  \ge \left\{ \begin{array}{l}
-1 + \sum_i V_i (d_i - 5 ) \\[4pt]
                                                  0
                                                \end{array}
                                                \right.
                                                  \,,
\ee
where the upper inequality results from the condition $N \le \sum_i V_i +1$.
As for the lower inequality, we made use of the fact that diagrams
made out of $m$ $C_4$-vertices  and $m+1$ insertions of $T^{(-1)}$ are
already included at LO. Thus, higher-order corrections $\propto C_4$
emerge from diagrams with at least two $C_4$-vertices separated by the
free Green's function, which start contributing at order $\sim M_{\rm lo}^0$.
This shows that all contributions to the amplitude beyond LO are perturbative. 

In Fig.~\ref{fig:diagrams_pc1}, we show the contributions to the
amplitude at next-to-leading order (NLO) and next-to-next-to-leading
order (NNLO). Evaluating the two order-$M_{\rm lo}^0$ diagrams, where
we have chosen to work with the operator $C_6 p'^3 p^3$, we obtain
the following contribution to the inverse $T$-matrix:
\be
\frac{4 \pi k^2}{m} \, \frac{T^{(0)} (k)}{\big[T^{(-1)} (k) \big]^2} =
\left[ - \frac{3 (\pi r + 4
      \mu_1)^2}{8 \pi \mu_3^3} + \frac{m C_6^R \mu_3^6}{9 \pi^3}
  \right] k^4 + \mathcal{O} (k^6)\,. 
  \ee
  Matching this expression to the first shape term in the ERE leads
  to\footnote{Here and in what follows, we made a choice to also keep
    higher-order contributions to various LECs as is a matter of convention.}
  \be
C_6^R (\mu_1, \mu_3) = \frac{9 \pi^2 [- 48 \mu_1^2 + \pi (3 \pi r^2 +
  8 v_2 \mu_3^3)]}{8 m \mu_3^9}
\; \sim \; \mathcal{O} (1)\,. 
\ee
Similarly, for the NNLO contribution, we find
 \bea
-  \frac{4 \pi k^2}{m} \left\{- \frac{T^{(1)} (k)}{\big[T^{(-1)} (k)
     \big]^2} +  \frac{\big[T^{(0)} (k)\big]^2}{\big[T^{(-1)} (k)
     \big]^3} \right\} &=&
\left[\frac{2 \delta C_4^{(3)} \mu _3^6 m}{9 \pi ^3}-\frac{3
   \left(4 \mu _1+\pi  r\right)}{2 a \mu _3^3}\right] k^2 
\ \\[4pt]
&+&
 \left[-\frac{9 \mu _1
   \left(4 \mu _1+\pi  r\right)}{2 a \mu _3^6}+\frac{\delta C_6^{(1)}  \mu _3^6 m}{9 \pi ^3}+\frac{4
   \delta C_4^{(3)}  \mu _1 \mu _3^3 m}{3 \pi ^3}+\frac{3 r \left(4 \mu _1+\pi  r\right)}{4 \mu
   _3^3}\right] k^4
\nonumber \\[4pt]
&-& \frac{i}{12 \mu _3^6}
\left[\frac{27 \pi  \left(4 \mu _1+\pi 
   r\right)}{a}-\frac{8 \delta C_4^{(3)} \mu _3^9 m}{\pi ^2}+18 \mu _3^3 \left(4 \mu _1+\pi 
 r\right)\right] k^5 + \mathcal{O} (k^6)\,.
\nonumber
\eea 
Matching this expression to the ERE leads to
\bea
  \label{DelC4}
\delta C_4^{(3)} (\mu_1, \mu_3) &=&\frac{27 \pi^3 (\pi r + 4 \mu_1)}{4
  a m \mu_3^9}
\,, \nonumber \\[4pt]
\delta C_6^{(1)} (\mu_1, \mu_3) &=&- \frac{27 \pi^3 (\pi r + 4 \mu_1)
  (6 \mu_1 + a r \mu_3^3)}{4 a m \mu_3^{12}}
\,.
\eea
Substituting these values back into the expression for the amplitude,
the NNLO result finally takes the form 
\bea
\label{FinalAmpl}
-\frac{4 \pi k^2}{m} \left\{ \frac{1}{T^{(-1)} (k)} -\frac{T^{(0)} (k)}{\big[T^{(-1)} (k)
     \big]^2} - \frac{T^{(1)} (k)}{\big[T^{(-1)} (k)
     \big]^2} +  \frac{\big[T^{(0)} (k)\big]^2}{\big[T^{(-1)} (k)
     \big]^3} \right\} &=&
 - \frac{1}{a} + \frac{1}{2} rk^2 - i k^3 + v_2 k^4 + i
 \frac{9 \pi (\pi r + 4 \mu_1 )}{4 a \mu_3^6} k^5 \nonumber \\
&+& \mathcal{O} (k^6)\,.
\eea
Notice that the last term in the first line of this equation violates unitarity but is of
order $\sim M_{\rm lo}^8$, which is beyond the accuracy of the NNLO approximation.  
Higher-order corrections to the amplitude can be calculated 
straightforwardly along the same lines and, for the case at hand, just
restore the ERE.  

As already pointed out above, we could have chosen the renormalization
conditions by setting $\mu_1 \sim M_{\rm lo}$, $\mu_3 \sim M_{\rm hi}$ as an
alternative to Eq.~(\ref{RenormCond}). Eq.~(\ref{TC2}) shows that in such an approach, the amplitude
$T_{C_2}$ is even stronger enhanced relative to NDA, namely   
$T_{C_2} \sim M_{\rm lo}^{-1}$. On the other hand, the subleading
interaction with $C_4^R \sim M_{\rm lo}$ is suppressed compared to NDA, reflecting directly the
fine tuned value of the effective range $r \sim M_{\rm lo}$. Regardless of these changes,
all diagrams in the first line of
Fig.~\ref{fig:diagrams_pc1} contribute at the same order ($\sim M_{\rm lo}^{-1}$) and must be
resummed to generate the LO amplitude $T^{(-1)}$. The perturbative
expansion of the amplitude has the same form as before, but the
corrections $\delta C_4$ and $\delta C_6$ are pushed to higher orders
and need not be taken into account at NNLO. 

\begin{figure}[t]
\includegraphics[width=\textwidth]{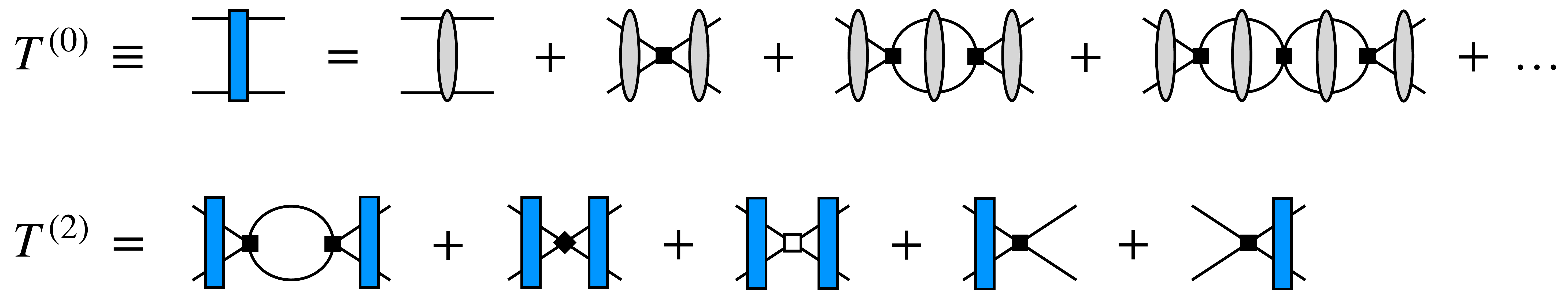}
\caption{Diagrams contributing to
  the P-wave scattering amplitude up to order $M_{\rm lo}^2$ for the scenario of
  Eq.~(\ref{pars}). Open rectangle denotes a correction to the $C_4$-vertex
  proportional to  $\delta C_4^{(2)}$ as explained in the text. For further notation,
  see Fig.~\ref{fig:diagrams_pc1}.}
\label{fig:diagrams_pc2} 
\end{figure}

The less fine-tuned scenario corresponding to Eq.~(\ref{pars}) can be
treated analogously. The LO contribution to the amplitude appears at
order $\sim \mathcal{O}(1)$ from the same set of diagrams as visualized
in Fig.~\ref{fig:diagrams_pc2}.
Notice that while the amplitude $T_{C_2}$ is enhanced by $\sim M_{\rm lo}^{-2}$ relative to
NDA as a result of the fine-tuned value of
$C_2^R$ in Eq.~(\ref{C2R}), the resummed contribution of diagrams
shown in the first line of Fig.~\ref{fig:diagrams_pc2}
is not enhanced any further since the value of $C_4^R$ is not fine-tuned. 
We further emphasize that differently to the EFT formulation with a
dimer field \cite{Bedaque:2003wa}, the LO amplitude is valid up-to-and-including
order $M_{\rm lo}$ and already incorporates the term $-i k^3$ in the
denominator stemming from the unitarity cut.  
For the renormalization conditions specified in Eq.~(\ref{RenormCond}),
corrections to the LO amplitude emerge from diagrams involving
higher-order vertices and insertions of $T^{(0)}$, where the power
counting expression (\ref{PowC1}) now takes the form 
\be
\label{PowC2}
n = 2 - 2N + \sum_i V_i (d_i - 2 )  \ge \left\{ \begin{array}{l} \sum_i V_i (d_i - 4 )\\[4pt]
                                                  2
                                                \end{array}
                                                \right.
                                                  \,.
\ee
Evaluating the diagrams shown in the second line of
Fig.~\ref{fig:diagrams_pc2}  with $C_2^R$ and  $C_4^R$
given in Eqs.~(\ref{C2R}), (\ref{C4leading}) and performing matching at the level
of the ERE, the LEC $C_6^R$ is found to be
\be
C_6^R (\mu_1, \mu_3) = \frac{9 \pi^2 \big[ - 36 \pi \mu_1 (\pi r + 4
  \mu_1) - 3 a (\pi r + 4 \mu_1)^2 \mu_3^3 + 8 \pi a v_2
  \mu_3^6\big]}{8a m \mu_3^{12}} \; \sim \; \mathcal{O} (1)\,,
\ee
while the expression for $\delta C_4^{(2)}$ coincides with that of
$\delta C_4^{(3)}$ in Eq.~(\ref{DelC4}).   
The resulting expression for the inverse of the amplitude $T^{(0)}(k) + T^{(2)}(k)$ is then
identical to the one given in Eq.~(\ref{FinalAmpl}).

While the doubly fine-tuned scenario of Eq.~(\ref{pc1}) only supports
wide shallow resonances with ${\rm Re} \, k_{\rm res}  \sim {\rm Im} \, k_{\rm res}  \sim M_{\rm lo}$,
where $k_{\rm res}$ denotes the location of the resonance pole,  less fine-tuned systems
described by Eq.~(\ref{pars}) may feature narrow resonances with
${\rm Re} \, k_{\rm res}  \sim M_{\rm lo}$,  ${\rm Im}\, k_{\rm res}  \sim M_{\rm lo}^2/M_{\rm hi}$.
For the near-resonance kinematics with $k \sim {\rm Re} \, k_{\rm res} $, the LO amplitude
$T^{(0)} (k)$ is enhanced and  scales as $\sim M_{\rm lo}^{-1}$ rather than $\sim M_{\rm lo}^{0}$.
In such a narrow kinematical region, the expansion of the amplitude actually coincides with that
shown in Fig.~\ref{fig:diagrams_pc1} apart from the contribution of
the $\delta C_4$-term, which appears already at NLO (i.e., in $T^{(0)} (k)$).  

It should be understood that a perturbative calculation of the
amplitude as demonstrated above is, strictly speaking, not necessary
for the case of separable interactions considered here, since the
LS equation  can be solved exactly for the potential
truncated at any order. In this way, one can obtain exact 
expressions for the renormalized LECs that incorporate all perturbative
corrections $\delta C_n^{(s)}$ discussed above. For example, solving the LS
equation for the potential given by the first two terms in
Eq.~(\ref{potPwave}), performing subtractive renormalization of the
amplitude with $\mu_5 = \mu_7 = 0$ and matching the LECs
to reproduce the scattering length and effective range leads to 
\be
\label{LECsRen}
C_2^R (\mu_3 )  = \frac{12\pi^2 }{m (3 \pi a^{-1}- 2  \mu_3^3 )}\,, \quad \quad
C_4^R (\mu_1, \mu_3 )  = \frac{9\pi^2  (\pi r + 4 \mu_1)}{m (3
  \pi a^{-1}- 2  \mu_3^3 )^2} \,,
\ee
in agreement with Eqs.~(\ref{C2R}), (\ref{C4leading}) and (\ref{DelC4}). 
We further emphasize that choosing different renormalization
conditions with $\mu_{\ge 5} \sim M_{\rm hi}$ would still result in the
same (perturbative) expansion of the scattering amplitude after
expressing the LECs in terms of
physical parameters (i.e., coefficients in the ERE). One would,
however, lose the manifest power counting for renormalized loop
diagrams written in terms of the LECs $C_n^R$ $C_n^{m, R}$, as they
would all appear to contribute at the same order. 

The proposed EFT formulation can be straightforwardly generalized to
resonant systems in partial waves with $l \ge 2$. Focussing
again on the cases in which the effective range function has no poles
for $k \sim M_{\rm lo}$, the strongest possible fine-tuning
corresponds to the first $l+1$ terms in the ERE being suppressed
compared to NDA and contributing at the same order as the unitary term,
i.e.~$a_l^{-1} \sim M_{\rm lo}^{2l+1}$, $r_l \sim M_{\rm lo}^{2l-1}$,
$v_{2, \, l} \sim M_{\rm lo}^{2l-3}$, $\ldots$. A self-consistent
Weinberg-like power counting scheme emerges by choosing the
subtraction scales according to $\mu_{2l+1} \sim M_{\rm hi}$, $\mu_{2l+3} \sim
\mu_{2l+5} \sim \ldots \sim  M_{\rm lo}$. Choosing the remaining
$\mu_i$'s of the order of the hard scale, i.e.~$\mu_1 \sim \mu_3 \sim
\ldots \sim \mu_{2l-1} \sim M_{\rm hi}$, will result in the renormalized
LECs that  all scale according to 
NDA.  On the other hand, choosing some of these scales of the order
$\sim M_{\rm lo}$ would lead to some of the  LECs accompanying the
interactions with $2l+2$ and more derivatives being suppressed.
For $\mu_1 \sim \mu_3 \sim \ldots \sim \mu_{2l-1} \sim M_{\rm lo}$,
the resulting scaling of the LECs
$C_{2l+2}^R \sim M_{\rm lo}^{2l-1}$,   $C_{2l+4}^R \sim M_{\rm lo}^{2l-3}$,
$\ldots$ $C_{4l}^R \sim M_{\rm lo}$ is in a one-to-one correspondence with the
scaling of the fine-tuned coefficients in the ERE. 

Last but not least, we emphasize that the proposed EFT formulation is by no
means restricted to the use of subtractive renormalization. Any
regularization scheme that provides sufficient flexibility to
incorporate the proper renormalization conditions is  
equally well suited for our purpose, and the results of the
calculations are, of course, independent on the choice of regulator. 
For example, one can apply dimensional regularization in the partial
wave basis (i.e., with the angular integrations being performed in
$d=4$ space-time dimensions) as discussed in Ref.~\cite{Phillips:1998uy}. 
The regularized integrals $J_{2s+1} (k)$ with $s \ge l$ are then given by
\bea
J_{2s+1} (k) = \int \frac{l^2 dl}{2 \pi^2} \frac{m l^{2s}}{k^2 - l^2 +
  i \epsilon} &\to & \left(\frac{\mu}{2} \right)^{4-d} \int
\frac{l^2 d^{d-3}l}{2 \pi^2}   \frac{m l^{2s}}{k^2 - l^2 +
  i \epsilon} \nonumber \\
&=& - \frac{m}{\pi} \left(\frac{\mu}{2}\right)^{4-d}k^{2s}
(-k^2 - i \epsilon)^{(d-3)/2} \; \Gamma \left( \frac{d-1}{2} \right)
\Gamma \left( \frac{3-d}{2} \right) \,.
\eea
While both the standard MS/$\overline{\text{MS}}$ scheme and the PDS
scheme of Ref.~\cite{Kaplan:1998tg} are too restrictive for the
considered fine-tuned systems,
one can introduce a generalized PDS approach, where poles in $d=3-2l$
dimensions,
\be
J_{2s+1} (k) \Big|_{\rm pole} = \frac{m k^{2 (s-l)} \mu^{2l+1}}{2^{2l}
  \pi \left[d- (3-2l)\right]} \,,
\ee
are subtracted from the analytically continued expressions
for the loop integrals in $d$ dimensions by taking into account the corresponding
counterterms (which are finite in $d=4$ dimensions). 
For the considered systems,
the resulting scheme represents a particular case of the more general
subtractive renormalization approach with all $\mu_i$ being set to zero
except for $\mu_{2l+1}$, whose value is related to the DR scale $\mu$
via
\be
\mu_{2l+1} = \left( 2^{1-2l} \pi \right)^{1/(2l+1)} \mu\,.
\ee
Alternatively, one can also choose to additionally
subtract poles in $3, \, 1, \, \ldots , \, 5-2l$ dimensions. 
In all cases, setting $\mu \sim M_{\rm hi}$ is necessary for the resulting EFT
to feature a consistent power counting scheme as described above.

\section{Wilsonian RG analysis}
\label{WRG}

In the previous section, we have discussed in detail the formulation of halo EFT
for resonant P-wave systems in terms of the renormalized potential. 
Below, we analyze the corresponding bare
potentials by means of the  Wilsonian RG flow equation following the
philosophy of Refs.~\cite{Birse:1998dk,Birse:2009my,Harada:2006cw} and discuss the
implications for the EFT.

\subsection{Wilsonian RG equation and fixed-point solutions for P-wave
  scattering}
\label{fixedpoints}

In Section~\ref{Halo_Nuclei}, we have presented a self-consistent formulation
of the nonrelativistic EFT for resonant two-body  P-wave scattering by
short-range forces. The key element of our consideration was the
knowledge of the general analytic structure of the on-shell scattering
amplitude $T(k)$ parametrized in terms of the ERE, which allowed us to identify
its expansion patterns for various scenarios and 
specify the appropriate renormalization conditions. The Wilsonian
RG approach we discuss below aims to achieve similar goals, but from a
different perspective and with no
reliance upon the ERE. Instead, various expansion regimes
corresponding to different physical situations
are identified by studying the RG flow in the 
parameter space describing generic (bare) short-range potentials and
analyzing perturbations around fixed-point solutions of the RG
equation. 

Following the philosophy of the Wilsonian RG analysis of Refs.~\cite{Birse:1998dk,Birse:2009my}, we 
study the evolution of a theory, specified by an 
energy-dependent potential, upon continuously integrating out
momentum modes above some cutoff scale $\Lambda$ while keeping
the off-shell scattering amplitude unchanged.   
The running of the potential with the cutoff can be inferred from
the LS equation for the off-shell $K$-matrix in the partial-wave basis
\be
K(p',p,k)=V(p',p, k, \Lambda )+  m \dashint_0^\Lambda \frac{l^2 d l}{2\pi^2} \,
\frac{V(p,l, k, \Lambda ) \;K(l,p',k)}{k^2-l^2}\,,
\label{eqLSUnderlying}
\ee
where the symbol $\dashint$ denotes the Cauchy principal value
integral. The real $K$-matrix is related to the $T$-matrix considered
in the previous sections via  $1/K(p',p,k) = {\rm Re} [1/T(p',p,k) ]$. 
Taking the derivative with respect to $\Lambda$ and using the LS
equation (\ref{eqLSUnderlying}), we arrive at the differential equation
\be
\label{PreRG}
\frac{\partial V}{\partial \Lambda} = \frac{m}{2 \pi^2} V(p', \Lambda,
k, \Lambda ) \frac{\Lambda^2}{ \Lambda^2-k^2} V(\Lambda, p,
k, \Lambda )\,.
\ee
The RG equation emerges by expressing all dimensionful quantities
in units of $\Lambda$ via $k = \hat k \Lambda$,   $p = \hat p \Lambda$ and  $p' = \hat
p' \Lambda$ and introducing the rescaled dimensionless potential
\be
\label{Rescaling}
\hat
V (\hat p', \hat p, \hat k, \Lambda) = \frac{m \Lambda}{2 \pi^2}
V ( p',  p,  k, \Lambda) \,,
\ee
defined in such a way that the factor of $m/(2 \pi^2)$ disappears from
the LS equation. Expressing Eq.~(\ref{PreRG}) in terms of the
rescaled quantities then yields the RG equation
\cite{Birse:1998dk,Birse:2009my}
\be
\label{RGEq}
\Lambda \frac{\partial \hat V}{\partial \Lambda} =
\hat p' \frac{ \partial \hat V}{\partial \hat p'} + \hat p \frac{\partial
 \hat  V}{\partial \hat p}
+ \hat k \frac{\partial \hat V}{\partial \hat k} + \hat V + \hat V
  (\hat p', 1, \hat k, \Lambda)  \frac{1}{1- \hat k^2} \hat V
  (1, \hat p, \hat k, \Lambda)  \,.
\ee
When lowering the cutoff towards $\Lambda \to 0$, the rescaled
potential $\hat V$ becomes cutoff independent once $\Lambda$ is pushed
well below all low-energy scales of the theory, i.e.~the potential flows
towards a fixed point solution  $\hat V (\hat p', \hat p, \hat k )$ that
describes a scale-invariant system. 
Notice that the RG equation always possesses a trivial
fixed point solution with $\hat V (\hat p', \hat p, \hat k ) = 0$
corresponding to the vanishing $K$-matrix.

Since we are interested here in  halo EFT with short-range
interactions only, we
can, without loss of generality, restrict ourselves to potentials of
separable type. Nontrivial fixed points can then be constructed
straightforwardly following the approach of Ref.~\cite{Birse:2015iea}. 
Specifically, consider rank-one separable potentials of the form
$\hat V (\hat p', \hat p, \hat k, \Lambda ) = \hat p' \hat p \, \hat \omega(\hat k, \Lambda )$
as relevant for the case of P-wave scattering.
A more general case of rank-two separable potentials is discussed in
Appendix~\ref{AppA}. 
The RG equation (\ref{RGEq}) then turns into an
ordinary differential equation for $\hat \omega (\hat k, \Lambda )$, which
becomes linear when expressed in terms of  $[\hat \omega (\hat k, \Lambda )]^{-1}$:
\be
\label{RGInverse}
\Lambda  \frac{\partial \hat \omega^{-1}}{\partial \Lambda}  = 
\hat k \frac{\partial \hat \omega^{-1}}{\partial \hat k} - 3 \hat \omega^{-1} -
\frac{1}{1 - \hat k^2}\,.
\ee
Integrating this equation for the fixed-point solution with $\partial
\hat \omega^{-1} / \partial \Lambda = 0$, subject to the boundary
condition that $\hat \omega (\hat k)$ is an analytic function of $\hat k^2$
for $\hat k \ll 1$, leads to
\be
\label{UnitaryFP}
\hat \omega_{\rm U} (\hat k) = \frac{-6}{2 + 6 \hat
  k^2 - 3 \hat k^3 \ln \frac{1 + \hat k}{1 - \hat k}}\,.
\ee
This fixed point is relevant for our considerations as it describes
P-wave systems in the unitary limit with $1/K(p',p,k) =0$. 

The trivial and the unitary fixed points $\hat V_{\rm T}= 0$ and
$\hat V_{\rm U}$, in order,  describe idealized situations we are not really
interested in. Rather, we want to describe  realistic systems that can be
approximated by perturbations about these idealized cases. For such
systems, the expansion patterns of the scattering amplitude in powers
of the ratio of the soft and hard scales can be determined by
analyzing perturbations about the fixed point solutions that scale
with definite powers of $\Lambda$ \cite{Birse:1998dk}. It is
sufficient for our purposes to study purely energy-dependent
perturbations of the form 
\be
\label{Perturb}
\hat \omega (\hat k, \Lambda ) = \hat \omega (\hat k ) +  \sum_\nu
C_\nu \, \Lambda^\nu \phi_\nu
(\hat k )\,,
\ee
where the $C_\nu$ are dimensionful coefficients while the functions $\phi_\nu (\hat
k )$ and the powers $\nu$ are to be determined. A more general case 
of momentum-dependent perturbations can be studied along the
lines of Ref.~\cite{Birse:1998dk}, but they generally appear to
contribute at higher orders. Solving the linearized RG equation
\be
\label{RGEqLin}
\nu \phi_\nu = 3 \phi_\nu + \hat k \frac{\partial \phi_\nu}{\hat k} +
2 \frac{\hat \omega (\hat k)}{1 - \hat k^2} \phi_\nu\,,
\ee
subject to the constraint that the perturbations are analytic
functions of $\hat k^2$ for  $\hat k \ll 1$,  one obtains
\bea
\label{PerturbResult}
\phi_\nu (\hat k ) &=& \hat k^{2n}\quad \text{with} \quad 2n = \nu - 3
= 0,\, 
2, \, 4, \, \ldots \,, \quad \text{for} \quad \hat \omega (\hat k) = \hat \omega_{\rm T}
(\hat k) \,, \nonumber \\
\phi_\nu (\hat k ) &=& \hat k^{2n} [ \hat \omega_{\rm U} (\hat k) ]^2 \quad
\text{with} \quad 2n = \nu +3  = 0, \, 
2, \, 4, \, \ldots \,, \quad \text{for} \quad \hat \omega (\hat k) = \hat \omega_{\rm U} (\hat k)\,.
\eea
Notice that for perturbations around nontrivial fixed points
such as $\hat \omega_{\rm U} (\hat k )$, one can, alternatively to the
linearized equation (\ref{RGEqLin}) for $\hat \omega (\hat k, \Lambda)$, use the linear
RG equation (\ref{RGInverse}) for $[\hat \omega (\hat k, \Lambda)]^{-1}$ \cite{Birse:2010fj} 
to obtain
\be
\label{InvOm}
\frac{1}{\hat \omega (\hat k, \Lambda)} = \frac{1}{\hat \omega_{\rm U} (\hat k
  )} - \sum_{n=0}^\infty C_{2n-3}' \Lambda^{2n - 3} \hat k^{2n}\,,
\ee
where $C_i' = C_i + \mathcal{O} (C^2 )$.   From the point of view of
the RG flow, the appearance of negative values of $\nu$ signals that
the corresponding fixed point is unstable.  For the unitary fixed
point $\hat \omega_{\rm U}$, one has two relevant directions corresponding
to $\nu= -3, \, -1$, see also Ref.~\cite{Harada:2007ua}, which bring the system away
from $\hat \omega_{\rm U}$ when lowering the cutoff $\Lambda$. 
Potentials that do not reside 
on the critical surfaces of nontrivial fixed points\footnote{Such critical surfaces correspond to
subspaces of the theory space, for which the
potentials are attracted to nontrivial fixed points in the limit
$\Lambda \to 0$.} flow  in the  $\Lambda \to 0$ limit
towards the stable trivial fixed point, which
possesses only irrelevant perturbations with $\nu > 0$.  
In this deep infrared (IR) regime,  the running
of the potential is, therefore, controlled by  the expansion around the trivial
fixed point. If all dimensionless parameters $\hat C_\nu \equiv
C_\nu/M_{\rm hi}^\nu$ that characterize the system are of order $\sim 1$
(as one would naturally expect since $M_{\rm hi}$ is the breakdown scale of
the derivative expansion), the perturbative
expansion of the potential around $V_{\rm T} = 0$ as defined in
Eqs.~(\ref{Perturb}), (\ref{PerturbResult}) also holds for $\Lambda \sim M_{\rm lo}$ and
for $M_{\rm lo} \ll \Lambda \ll M_{\rm hi}$.
Such situations describe weakly interacting ``natural'' P-wave systems, and the
scattering amplitude can be calculated perturbatively.

An alternative expansion of the potential around the unitary fixed point
\be
\label{PotPerturbU}
V (p', p, k, \Lambda ) = \frac{2 \pi^2}{m \Lambda} \, \frac{p' p}{\Lambda^2}
\, \left\{\hat \omega_{\rm U}(k/\Lambda) +
  \left( \frac{M_{\rm hi}}{\Lambda} \right)^3 \sum_{n=0}^\infty \hat
  C_{2n-3}  \left( \frac{k}{M_{\rm hi}} \right)^{2n} \big[  \hat \omega_{\rm U}(k/\Lambda) \big]^2
  \right\}
  \ee
can be interpreted most easily by noticing the one-to-one
correspondence of the inverse potential with $[\hat {\fet \omega}(k/\Lambda, \Lambda )]^{-1}$
fulfilling Eq.~(\ref{InvOm}) 
with the ERE \cite{Birse:1998dk}, which follows immediately if the LS
equation is written  in the operator form as $T^{-1} = V^{-1} + G_0$: 
\be
\label{RGERE}
\frac{k^2}{K(k)} = - \frac{m}{4 \pi} \left( - \frac{1}{a} +
  \frac{1}{2} r k^2 + v_2 k^4 + \ldots \right) = - \frac{m M_{\rm
    hi}^3}{2 \pi^2} \sum_{n=0}^\infty \hat C_{2n-3}' \left(
  \frac{k}{M_{\rm hi}} \right)^{2n}\,.
\ee
The parameters $\hat C_i '$ can thus be expressed in terms of the
coefficients in the ERE:
\be
\hat C_{-3}' = - \frac{\pi}{2 a M_{\rm hi}^3}, \quad \quad
\hat C_{-1}' = \frac{\pi r}{4 M_{\rm hi}}, \quad \quad
\hat C_{1}' = \frac{\pi v_2 M_{\rm hi}}{2}, \quad \quad
\ldots \,.
\ee
For an expansion around the unitary fixed point to be valid,
the perturbations in Eq.~(\ref{PotPerturbU}) must be suppressed compared to the
LO term corresponding to the unitary fixed point.
For ``natural'' systems with all $\hat C_\nu \sim 1$,
the irrelevant perturbations $\propto \hat C_\nu$ with $\nu > 0$ are
indeed small, but the relevant ones $\propto \hat C_{-3}$ and $\propto \hat C_{-1}$
are much larger than the first term in the curly
brackets. Indeed, as we discussed above, ``natural'' systems
rather correspond to the expansion around the trivial fixed point.  
However, for systems with the strength of the relevant
perturbations being fine-tuned to unnaturally small values $a^{-1} \sim M_{\rm lo}^3$  and $r\sim M_{\rm lo}$
as given in Eq.~(\ref{pc1}), all perturbations in
Eq.~(\ref{PotPerturbU}) are indeed suppressed compared to the LO term for
$M_{\rm lo} \ll \Lambda \ll M_{\rm hi}$.  Such systems reside close to the
critical surface of the unitary fixed point, thus being attracted to $\hat V_{\rm U}$
for not too small values of $\Lambda$. For $\Lambda \sim M_{\rm lo}$, the two relevant
perturbations become large and must be resummed. For even smaller cutoff values, these increasing
perturbations drive the system towards the trivial fixed point.
Notice that for the less fine-tuned scenario in Eq.~(\ref{pars}), the
relevant perturbations are not suppressed for any $\Lambda \ll M_{\rm hi}$.
One, therefore, cannot expect such systems to be described by
the expansion around the unitary fixed point.  

To summarize, the Wilsonian RG analysis deals with the behavior of
generic bare potentials in the IR regime with $\Lambda \ll M_{\rm hi}$.
It allows one to identify certain expansion patterns of the
scattering amplitude in powers of $M_{\rm lo}/M_{\rm hi}$ from
analyzing the running of the potentials near the fixed points of
the RG equation. For two-body scattering of nonrelativistic particles
interacting via short-range forces we are interested in here, the RG
analysis can be carried out analytically, yielding, however, essentially an alternative
derivation of the ERE for the scattering amplitude. More interesting
and nontrivial examples include applications of the RG analysis to
systems interacting with both long- and short-range forces as
relevant for chiral EFT for nuclear systems, see
Refs.~\cite{Barford:2002je,Birse:2005um,Birse:2007sx,Ando:2008jb,Birse:2010jr,Harada:2010ba,Harada:2013hwa,Harada:2013zya}
for some work along this line.

\subsection{Implications for the EFT}
\label{FPsHPWs}

While our considerations in the previous section in terms of the bare
potentials have been rather general, they may appear unrelated
to the subtractively renormalized formulation of halo EFT
considered in Section~\ref{Halo_Nuclei}. The purpose of this section is to
unmask the relationship between the two approaches and to address
implications of the RG analysis to the power counting of the halo EFT. 

To establish a connection between the two approaches, we 
consider the bare energy-dependent potential $V(p', p, k)$ defined in
Eq.~(\ref{BarePot}) and corresponding to
the subtractively renormalized potential $V^R (p', p) = C_2^R  p' p + C_4^R  p' p (p'^2 + p^2)$.
To verify that it indeed complies with the expansions discussed in the previous section, 
it is more convenient to rewrite it in terms of the physical parameters $a$
and $r$ instead of $C_2^R$ and $C_4^R$. The resulting bare potential 
defines a family of
physical systems characterized by the parameters $a$, $r$, $\mu_1$ and
$\mu_3$. From the two remaining scales, $\mu_7$ is, in fact, a
redundant parameter since the dependence of the amplitude on $\mu_7$ is completely
eliminated by the running of the LEC $C_2^R$. Consequently, the
potential does not depend on $\mu_7$ after being expressed in terms of
$a$ and $r$. The scale $\mu_5$ does not enter the
expression for the on-shell amplitude, cf.~Eq.~(\ref{ERFRen}), but affects
the off-shell behavior of the potential and scattering amplitude. For
the sake of definiteness, we fix the off-shell behavior of the potential
by choosing $\mu_5  = 0$ to obtain
\bea
\label{PotModel1}
V &=& - \frac{20 \pi ^2 p p'}{m} \bigg[840 a \left(2 a \mu _3^3-3 \pi \right)^3-3 a^2 \left(4 \mu _1+\pi 
  r\right) \bigg(
210 \left(3 \pi -2 a \mu _3^3\right)^2
   \left(p^2+p'^2\right) +
\frac{9}{2} a^2 \left(4 \mu _1+\pi  r\right) \bigg\{-35
\left(k^2-p^2\right) \nonumber \\
     &\times &    \left(k^2-p'^2\right) 
     \bigg[
     3 k^3 \ln \frac{\Lambda +k}{\Lambda
   -k} -6 k^2 (\Lambda - \mu _1) -2 (\Lambda ^3 - \mu _3^3)\bigg]+42 \Lambda ^5
 \left(k^2-p^2-p'^2\right)+30 \Lambda ^7\bigg\}\bigg)\bigg]
 \nonumber \\
     &\times & 
\bigg[
42 a^2 \left(4 \mu _1+\pi 
   r\right) \left[5 k^2 (\Lambda ^3 - \mu _3^3)  +15 k^4 (\Lambda
   -\mu_1 )+3 \Lambda ^5\right]
   \left[9 \pi  \left(a^2 \Lambda ^5 r-20 \pi \right)+4 a \left(9 a \Lambda ^5 \mu _1-20 a \mu _3^6+60
   \pi  \mu _3^3\right) \right] \nonumber \\
&-&50 a \left[3 k^2 (\Lambda - \mu _1) +\Lambda ^3-\mu _3^3\right] \left\{27 \pi ^2 \left(a^3 \Lambda ^7
   r^2+56 \pi \right)+8 a \left[27 a^2 \Lambda ^7 \mu _1 \left(2 \mu _1+\pi 
   r\right)-56 a^2 \mu _3^9+252 \pi  a \mu _3^6-378 \pi ^2 \mu
 _3^3\right]\right\}\nonumber \\
&-&15 a k^3 \ln \frac{\Lambda + k
   }{\Lambda -k} \Big(8 a \left\{27 \mu _1 \left[2 a^2 \Lambda ^5 \mu _1 \left(7 k^2-5 \Lambda
   ^2\right)+\pi  k^2 \left(7 a^2 \Lambda ^5 r-70 \pi \right)-5 \pi  a^2 \Lambda ^7 r\right]+280 a^2 \mu
   _3^9 \right.  \nonumber \\
&-& \left.  210 a \mu _3^6 \left[4 a k^2 \mu _1+\pi  \left(a k^2 r+6\right)\right]+630 \pi  \mu _3^3 \left[4
   a k^2 \mu _1+\pi  \left(a k^2 r+3\right)\right]\right\}-27 \pi ^2 \left[a^3 \Lambda ^5 r^2 \left(5
   \Lambda ^2-7 k^2\right)
\right. \nonumber \\
&+& \left. 140 \pi  \left(a k^2
    r+2\right)\right]
\Big)
+1400 \left(3 \pi -2 a \mu _3^3\right)^4
\bigg]^{-1}\,.
\eea
Notice that since we want the corresponding rescaled potential to fulfill the RG
equation, we have taken the limit $\Lambda \to \infty$ for the
renormalized integrals  $J_n^R (k, \mu_i)$ in Eq.~(\ref{BarePot}) to obtain 
the above expression. 
When inserted into the LS equation regularized with a sharp cutoff,
the above potential yields the $\Lambda$-independent off-shell amplitude
that reproduces Eq.~(\ref{ERFRen}) in the on-shell 
limit. It, therefore,  fulfills Eq.~(\ref{PreRG})
and may serve as a specific  example of generic bare
potentials considered in the previous section.

It is now instructive to
expand this potential in the ratio of the soft and hard scales
as defined in the RG analysis of Section~\ref{fixedpoints}. Specifically, we
assign  $\Lambda \sim k \sim p \sim p' \sim M_{\rm lo}$ and choose
$\mu_3 \sim M_{\rm hi}$ to comply with the considered physical
scenario by ensuring the absence of low-lying poles in
the inverse on-shell $K$-matrix, see the discussion in Section~\ref{Halo_Nuclei}.
As for the remaining scale $\mu_1$,
we consider here the case of $\mu_1 \sim M_{\rm hi}$, which allows one
to simulate the correction to the potential needed to reproduce the
shape parameter $v_2$ by tuning $\mu_1$.    
\begin{itemize}
\item Consider first the doubly fine-tuned scenario of Eq.~(\ref{pc1}) with
$a^{-1}\sim M_{\rm lo}^3$, $r \sim M_{\rm lo}$. 
Expanding the bare potential in Eq.~(\ref{PotModel1}), rescaled
according to Eq.~(\ref{Rescaling}), in powers of $\epsilon \equiv M_{\rm lo}/M_{\rm hi}$, one obtains
\be
\label{ExpExample1}
\hat V(\hat p', \hat p, \hat k, \Lambda )  = 
\frac{12 \, \hat p' \, \hat p}{-4 +  \frac{6 \pi}{a \Lambda^3}  - 3
  \hat k^2
\big(4 + \frac{\pi r}{\Lambda}
  \big) 
  +6 \hat k^3 \ln \frac{1+ \hat k}{1 -
    \hat k}} +
\frac{432\, \hat k^4 \, \hat p' \, \hat p \,  \mu_1^2 \, \Lambda}{\Big[ -4 +
  \frac{6 \pi}{a \Lambda^3}  - 3 \hat k^2 \big(4 + \frac{\pi r}{\Lambda}
  \big) +6 \hat k^3 \ln \frac{1+ \hat k}{1 -
    \hat k}\Big]^2  \mu_3^3}  + \mathcal{O}
(\epsilon^2 )\,.
\ee
In agreement with the considerations of Section~\ref{fixedpoints},
cf.~the second line in Eq.~(\ref{PerturbResult}),
the potential is described in terms of the expansion around the
unitary  fixed point with resummed corrections stemming from the relevant
perturbations $\propto a^{-1},  r$. The LO term $\hat V^{(0)}$ leads to the
effective range approximation $k^3 \cot \delta = - a^{-1} + r k^2/2$, 
while the order-$\epsilon$ correction $\hat V^{(1)}$ generates the first
shape term in the ERE if one chooses $\mu_1^2 = \pi v_2 \mu_3^3/6$ as
mentioned above. Notice that the LO term corresponds to the
theory that parametrizes the renormalized trajectories connecting
the unitary and trivial fixed points, see Appendix~\ref{AppA} for more
details.
\item
For the less fine-tuned scenario of Eq.~(\ref{pars}) with
$a^{-1} \sim M_{\rm lo}^2 M_{\rm hi}$ and $r \sim M_{\rm hi}$, the expansion of
the potential takes the form
\be
\label{ExpExample2}  
\hat V(\hat p', \hat p, \hat k, \Lambda )  = \frac{4 \hat p ' \hat
  p}{\pi \left( \frac{2}{a \Lambda^3} -  \frac{r}{\Lambda} \hat
  k^2 \right) } +
\frac{8 \hat p ' \hat p}{3 \pi^2
\left( \frac{2}{a \Lambda^3} -  \frac{r}{\Lambda} \hat
  k^2 \right)^2} \left( 2 + 6 \hat k^2 - 3 \hat k^3 \ln \frac{1+ \hat k}{1 -
    \hat k} \right) + \mathcal{O}
(\epsilon^3 )\,,
\ee
where the first and second terms contribute at orders $\epsilon$ and
$\epsilon^2$, respectively.
As expected from the general arguments of the previous section, this
situation corresponds to the expansion around the trivial fixed point with
resummed contributions from the scattering length and effective range. 
In this case, the LO potential $\hat V^{(1)}$ yields
\be
k^3 \cot \delta = \left( - \frac{1}{a} - \frac{2 \Lambda^3}{3 \pi}
\right) +  \left(\frac{r}{2} - \frac{2 \Lambda}{\pi} \right) k^2 +
\frac{2k^4}{\pi \Lambda} + \ldots\,,
\ee
thus indeed providing the LO contribution to the ERE, accompanied with
higher-order contributions. 
\end{itemize}  

One may now raise the question of what the expansion of the potential
would correspond to if all subtraction scales, including $\mu_3$, would have been chosen
of the order of the soft scale in the problem, such that the
scattering amplitude would feature low-lying zero(s).  As shown in Appendix~\ref{AppA},
the resulting resonant systems are described by expansions
around different unstable fixed points.

The above examples show that for nonrelativistic systems with a clear scale
separation, low-energy physics can be systematically described
by expanding the bare potential around fixed point solutions of the
RG equation. This method utilizes the same kind of expansion in powers
of $M_{\rm lo}/M_{\rm hi}$ as the corresponding EFT, and it has proven
to be particularly useful 
for analyzing universality aspects of strongly interacting
systems, see Ref.~\cite{Braaten:2004rn,Naidon:2016dpf} for review articles. In spite of the
similarities, the RG approach outlined above does not
directly translate into the EFT program in the way it is usually formulated,
which relies on (local) effective
Lagrangians and typically requires choosing $\Lambda \sim M_{\rm hi}$
to exploit the full predictive power. The
scattering length and effective range entering the bare potential in
Eq.~(\ref{PotModel1}), which obeys a well-defined expansion around the unitary fixed point
as given in Eq.~(\ref{ExpExample1}), are, in fact,  complicated nonlinear functions of the
parameters entering the effective Lagrangian, i.e.~of the
renormalized LECs $C_2^R (\mu_i)$ and $C_4^R (\mu_i)$, and the
expansion pattern of the renormalized potential and thus also of the
scattering amplitude depends crucially on
the choice of the scales $\mu_i$ (i.e.~on the renormalization
conditions), see Section~\ref{Halo_Nuclei} for details, which play a
role similar to  the floating cutoff $\Lambda$ in the Wilsonian RG
analysis. 
For S-wave systems near the unitary limit, all renormalization scale(s)
can be pushed down to $\mu_i \sim M_{\rm lo}$ as done in the KSW
approach, and the correspondence between the Wilsonian and  Gell-Mann
and Low RG  approaches becomes evident.  The
scaling behavior of the perturbations in the bare potential, expanded
around the unitary fixed point, then translates into the
scaling of the renormalized LECs in the KSW approach. The Wilsonian RG analysis
thus provides an alternative derivation of the KSW power counting. 
On the other hand, for resonant P-wave systems we are interested in
here, with the coefficients in the ERE scaling according to Eqs.~(\ref{pc1}) and (\ref{pars}),
choosing $\mu_3 \sim M_{\rm lo}$ corresponds, as already mentioned above, to a different
class of theories, see Appendix~\ref{AppA} for details. Thus, no KSW-like power counting scheme
with the LECs  $C_6^R$,  $C_8^R$, $\ldots$ being enhanced 
compared to their NDA scaling by the factor of $M_{\rm lo}^{-6}$,  as
suggested by Eq.~(\ref{PotPerturbU}), can be formulated for the case at hand.   
The contributions of these operators to the amplitude are, of course,
still enhanced for resonant systems  regardless of the choice of the
renormalization conditions (as follows from both the ERE and the Wilsonian
RG analysis). In the Weinberg-like power counting scheme formulated in Section~\ref{Halo_Nuclei},
all LECs scale according to NDA and the large anomalous
canonical dimensions of the corresponding operators are generated
through the choice of the renormalization conditions ($\mu_3 \sim M_{\rm hi}$),
see also Ref.~\cite{Harada:2006cw} for a related discussion. 

Last but not least, we emphasize that the essential ingredient of the
Wilsonian RG method outlined above is its restriction to the IR
regime with $\Lambda \ll M_{\rm hi}$, where the representation of the
effective potential in terms of the expansion in powers of momenta is valid. 
It cannot provide a systematic power counting for the bare potential
if the cutoff parameter is taken beyond the hard scale of the problem.
This is further illustrated in Appendix~\ref{ToyModelRG},
where the exact RG trajectory for a toy-model S-wave
potential with long-range interaction is calculated
numerically. Generally, moving against the RG  flow by increasing $\Lambda$   
beyond the hard scales of the problem, without at the same time taking into account the 
corresponding new degrees of freedom, as it is done in the lcRG-invariant approach,
is a dangerous endeavor. It typically leads to complex values of the
potential when written in terms of the LECs or brings it to
infinity for $\Lambda \to \infty$ (unless the theory lies on a
critical surface of some nontrivial fixed point).

\section{Summary and conclusions}
\label{sec:summ}
In this paper we have revisited the problem of renormalization in
low-energy EFTs of nuclear interactions on the example of 
resonant P-wave scattering. Following Refs.~\cite{Bertulani:2002sz,Bedaque:2003wa}, we
focused here on the fine-tuned scenarios with the coefficients in the
effective range expansion scaling according to Eqs.~(\ref{pc1}) or
(\ref{pars}) and leading to the appearance of shallow bound, virtual or
resonance states. While such resonant systems have already
been extensively studied using EFT formulations with auxiliary
dimer fields
\cite{Bertulani:2002sz,Bedaque:2003wa,Gelman:2009be,Alhakami:2017ntb,Schmidt:2018vvl,Ji:2014wta,Ryberg:2017tpv,Soto:2007pg}, see Ref.~\cite{Hammer:2017tjm} for a
review article, we have employed here the
effective Lagrangian written solely in terms of contact interactions.
Our main findings are summarized below.
\begin{itemize}
\item
We started with applying the lcRG-invariant approach of
Refs.~\cite{Hammer:2019poc,Valderrama:2016koj,Habashi:2020qgw}
to resonant P-wave scattering in Section~\ref{Halo_Nuclei_Literature}. The presence of shallow
states demands resummation of the contact
interactions $C_2 p' p$ and   $C_4 p' p (p'^2 + p^2)$ when calculating the
scattering amplitude. However, solving the Lippmann-Schwinger equation and
expressing the bare LECs  $C_2 (\Lambda )$ and $C_4 (\Lambda )$ in
terms of the scattering length and effective range, we
found no solutions in terms of real LECs compatible with either of Eqs.~(\ref{pc1}) and (\ref{pars}) if the
cutoff is taken well beyond the hard scale in the problem, $\Lambda \gg M_{\rm hi}$.
This result is in agreement with the causality bounds
derived in Ref.~\cite{Hammer:2010fw}, but it appears to contradict the
conclusions obtained using the EFT with auxiliary
dimer fields \cite{Bertulani:2002sz,Bedaque:2003wa,Hammer:2017tjm}.
Indeed, keeping $\Lambda \sim M_{\rm  hi}$ shows that at least the less fine-tuned
scenario of Eq.~(\ref{pars}) is easily realizable in terms of a simple
quantum mechanical model, while it cannot be accommodated by the lcRG-invariant
approach.
\item
The above issue with the lcRG-invariant approach can be traced back to the
inconsistent (from the EFT point of view) renormalization of the LS equation with perturbatively
non-renormalizable potentials, which requires the inclusion of 
an infinite number of counterterms, see
e.g.~Ref.~\cite{Epelbaum:2018zli}. As repeatedly pointed out 
in Refs.~\cite{Lepage:1997cs,Epelbaum:2009sd,Epelbaum:2017byx,Epelbaum:2018zli,Epelbaum:2020maf},
arbitrarily large cutoff values can be employed in a way compatible with the principles of EFT only 
after all UV divergences, generated by iterations of the LS
equation, are removed. In Section~\ref{Halo_Nuclei}, we have shown how to
consistently renormalize the scattering amplitude for resonant
P-wave scattering in halo EFT with no auxiliary fields using a
subtractive scheme and utilizing the usual  QFT renormalization
technique to all orders in the loop expansion. A separable
form of the underlying effective potential admits a closed-form
expression for the (infinite set of) counterterms needed to absorb all divergences in
the LS equation as given in Eq.~(\ref{BarePot}). The resulting
scattering amplitude is finite in the limit $\Lambda \to \infty$, both
perturbatively (i.e., at any order in the loop expansion) and
non-perturbatively. A self-consistent power counting scheme is
obtained by choosing the subtraction scales according to
$\mu_3 \sim M_{\rm hi}$, $\mu_5 \sim \mu_7 \sim M_{\rm lo}$.  
These renormalization conditions ensure that (i) all renormalized LECs
scale according to NDA, (ii) the renormalized contributions of
diagrams obey manifest power counting, and their EFT order can be
determined a priori using the power counting formulas (\ref{PowC1}) and
(\ref{PowC2}) for the doubly and singly fine-tuned scenarios of
Eqs.~(\ref{pc1}) and (\ref{pars}), respectively, (iii) the
renormalized LECs $C_2^R$ and $C_4^R$ can be expressed in terms of $a$
and $r$ regardless of their actual values\footnote{Notice that
  contrary to what is claimed in Refs.~\cite{Habashi:2020qgw,Habashi:2020ofb}, renormalization by
  itself imposes no constraints on the relative sizes and signs of the scattering
  length and the effective range. Similarly to the EFT formulation
  with auxiliary fields \cite{Bertulani:2002sz,Bedaque:2003wa}, the
  framework we present here simply
  leads to the most general parametrization of the scattering amplitude
  compatible with the principles underlying its construction as formulated in
  Weinberg's theorem \cite{Weinberg:1978kz,Weinberg:1996kw}.
  It does, in particular, not guarantee the
  absence of unphysical poles on the upper half plane of the complex
  momentum plane. This feature follows from analytic properties of the scattering
  amplitude for certain classes of energy-independent potentials
  \cite{Newton}, but it does not hold for energy-dependent interactions like the
  one in Eq.~(\ref{BarePot}). The absence
  of unphysical poles of the $S$-matrix in the lcRG-invariant analysis of
  resonant S-wave systems in Ref.~\cite{Habashi:2020qgw}, the feature
  that has been
  attributed to renormalization in that paper, is simply a consequence of
  their renormalization procedure being realized entirely within a quantum mechanical
  framework with energy-independent interactions. For the EFT formulation
  we use here, the parameter sets leading to spurious poles of the
  $S$-matrix,  i.e.~the corresponding combinations of $a$, $r$ and $v_i$, should be
  regarded as unphysical and discarded.} in a close analogy with the
EFT formulations of Refs.~\cite{Bertulani:2002sz,Bedaque:2003wa},  (iv) the residual dependence of
the amplitude on the subtraction scales $\mu_i$ is beyond the
actual order of the calculation and (v) the EFT expansion is compatible with the
required scenarios in Eqs.~(\ref{pc1}) and (\ref{pars}). The choice of
the scale $\mu_3 \sim M_{\rm hi}$ is dictated by the need to avoid the
appearance of low-lying amplitude zeros to comply with the assumed  
scaling behaviors in Eqs.~(\ref{pc1}) and (\ref{pars}). It is,
therefore, not possible to formulate a KSW-like power counting scheme
for the considered systems, where all subtraction scales would be chosen of
the order of $M_{\rm lo}$ and the enhancement of the resummed LO contribution
to the amplitude would emerge from the enhancement of the individual
diagrams through the enhanced renormalized LECs.

The EFT we propose is not
restricted to P-waves and can be straightforwardly generalized to
describe resonant systems with any value of the orbital angular
momentum. It also permits the use of dimensional regularization,
supplied with an appropriate subtraction scheme that allows sufficient flexibility
to implement the proper renormalization conditions, such as e.g.~the
generalized PDS scheme. This feature might be particularly beneficial for
applications to halo systems in the presence of external electroweak
probes.
\item
  Next, we have performed a Wilsonian RG analysis of P-wave
  scattering in Section~\ref{WRG} following the philosophy of Refs.~\cite{Birse:1998dk,Birse:2009my},
  see also
  Refs.~\cite{Harada:2006cw,Harada:2007ua} for a closely related approach. Our main motivation
  here was to clarify the relationship between this powerful 
  method, formulated in terms of bare potentials, 
  and the subtractively renormalized halo EFT framework developed
  in Section~\ref{Halo_Nuclei}. The key
  ingredient of the Wilsonian RG analysis is the search for fixed point
  solutions of the RG equation (\ref{RGEq}). In addition to the trivial fixed point
  that describes non-interacting systems, the unitary fixed
  point in Eq.~(\ref{UnitaryFP}) plays an important role for doubly fine-tuned
  systems specified in Eq.~(\ref{pc1}). This unstable fixed point
  describes scale-free P-wave systems with $a^{-1} \to 0$ and $r \to 0$ and has two
  relevant directions \cite{Harada:2007ua}. Once the floating cutoff $\Lambda$ is
  lowered well below the hard scale $M_{\rm hi}$, so that the
  expansion of the potential in terms of contact  interactions is
  valid, all theories describing doubly  fine-tuned systems in
  Eq.~(\ref{pc1}) get attracted by the unitary fixed point when  $M_{\rm lo} \lesssim
  \Lambda \ll M_{\rm hi}$. This allows one to identify a systematic
  and universal expansion of the scattering amplitude for such fine-tuned systems by
  analyzing the scaling of perturbations around the fixed point for
  $\Lambda \sim M_{\rm lo}$. For the case at hand, the Wilsonian RG
  analysis merely provides an alternative derivation of the
  ERE, cf.~Eq.~(\ref{RGERE}). It also implies that the contributions of the shape-terms to
  the scattering amplitude for doubly fine-tuned systems are enhanced by $M_{\rm lo}^{-6}$ as
  compared with NDA, see Eq.~(\ref{PotPerturbU}). This is, of course, in agreement with the ERE and,
  therefore, also with the EFT formulated in Section~\ref{Halo_Nuclei} as
  visualized in Fig.~\ref{fig:diagrams_pc1}.  On the other hand, the behavior of singly
  fine-tuned systems specified in Eq.~(\ref{pars}) is not expected to
  be  governed by the  expansion around the unitary fixed point.

  To further demonstrate the close relationship between the two approaches, we
  have considered the potential in Eq.~(\ref{BarePot}), which includes the resummed
  contributions of the counterterms in the subtractively renormalized
  EFT framework. After taking the limit $\Lambda \to \infty$
  in the renormalized UV-convergent integrals $J_n^R$, the resulting
  rescaled bare potential fulfills the RG equation (\ref{RGEq}). We have
  explicitly verified that the RG flow of this potential 
  indeed coincides with the expansion around the unitary fixed point
  for $\Lambda \sim M_{\rm lo}$, provided $\mu_3$ is chosen of the
  order $\sim M_{\rm hi}$ to
  comply with the conditions of Eq.~(\ref{pc1}). For singly
  fine-tuned systems specified in Eq.~(\ref{pars}), the expansion of
  the potential in powers of $M_{\rm lo}/M_{\rm hi}$ for $\Lambda \sim M_{\rm lo}$
  is found to coincide with that around the trivial fixed point with
  resummed corrections $\propto a^{-1}, \, r$. The general RG flow of 
  rank-two separable potentials like the one in Eq.~(\ref{PotModel1}) is
  discussed in Appendix~\ref{AppA} and shown to exhibit a rather rich
  structure.

  Last but not least, we emphasize that  taking $\Lambda \sim M_{\rm hi} $
  or larger is not compatible with the
  systematics underlying approximate expansions of the bare  
  potential within the Wilsonian RG analysis. To illustrate this point, we
  compared in Appendix~\ref{ToyModelRG} the exact RG flow for a toy-model
  S-wave potential, featuring 
  a long-range interaction, to the approximate result obtained using 
  the lcRG-invariant approach. Fixing one available parameter of the LO contact
  interaction as a function of $\Lambda$ from the 
  phase shift at some fixed energy, the resulting low-energy phase
  shifts are found to show very mild cutoff-dependence for $\Lambda\gg M_{\rm hi}$,
  thus (approximately) satisfying the condition of the RG
  invariance of the lcRG-invariant approach. However, the obtained
  limit-cycle-like $\Lambda$-dependence of the LO potential disagrees
  with the smooth RG flow behavior of the underlying model, a
  result that might have been expected given that the LO
  approximation to the bare potential is only valid for $\Lambda$ below $M_{\rm hi}$.    
\end{itemize}

Previous halo EFT studies of resonant systems in P- and higher partial waves
made use of the formulations with auxiliary dimer fields
\cite{Bertulani:2002sz,Bedaque:2003wa,Gelman:2009be,Alhakami:2017ntb,Hammer:2017tjm},
which are usually claimed to be introduced for convenience, see
e.g.~Ref.~\cite{Hammer:2019poc}. In this paper we have explicitly shown that the
EFT formulations with and without dimer fields are indeed equivalent.   
A remarkable aspect of this equivalence is that all diagrams contributing to 
the LO scattering amplitude in halo EFT with auxiliary dimer fields are
renormalizable, since all divergences from dressing the dimeron propagator
can be absorbed into its residual mass and the particle-dimeron
coupling constant. In contrast, the effective potential involving
contact interactions in the formulation without auxiliary fields is
not renormalizable in the usual sense. A proper renormalization of the
scattering amplitude, therefore, requires taking into account
contributions of an infinite number of counterterms. This unavoidably
introduces a dependence on the subtraction scales in the renormalized
amplitude, which reflects the freedom in choosing the finite pieces of
the corresponding coupling constants and can be kept to be of a higher
order by using the appropriate renormalization conditions as discussed in
Section~\ref{Halo_Nuclei}.  The resulting subtractively renormalized EFT is
indeed equivalent to halo EFT with auxiliary fields. On the other
hand, we have shown that these two EFT formulations are {\it not}
equivalent to the lcRG-invariant approach of Refs.~\cite{Nogga:2005hy,Hammer:2019poc,Valderrama:2016koj,Habashi:2020qgw}
if the requirement of $\Lambda \gg M_{\rm hi}$ is to be taken seriously.

\acknowledgments
This work was supported in part by BMBF (Grant No. 05P18PCFP1), by DFG and NSFC through funds provided to the
Sino-German CRC 110 ``Symmetries and the Emergence of Structure in QCD'' (NSFC
Grant No.~11621131001, Project-ID 196253076 - TRR 110), by Collaborative Research Center ``The Low-Energy
Frontier of the Standard Model'' (DFG, Project No. 204404729 - SFB 1044), by the Cluster of
Excellence ``Precision Physics, Fundamental Interactions, and Structure of Matter'' (PRISMA$^+$, EXC 2118/1)
within the German Excellence Strategy (Project ID 39083149),  
by the  Georgian Shota Rustaveli National Science Foundation (Grant No. FR17-354), by Volkswagen\-Stiftung
(Grant No. 93562), by the CAS President's International Fellowship Initiative (PIFI) (Grant No.~2018DM0034)
and by the EU (STRONG2020).

\appendix

\section{RG flow for rank-two separable P-wave potentials}\label{AppA}

The purpose of this appendix is to provide a detailed discussion
of the RG invariant bare potential of Section~\ref{FPsHPWs} and its
interpretation from the point of view of the RG flow. To this aim, we consider a more
general class of energy-dependent potentials as compared to our
considerations in Section~\ref{fixedpoints} of a rank-two separable form:
\be
\label{PotBareRank2}
V (p', p, k, \Lambda) = \fet \chi^T (p' ) \, \fet \omega (k , \Lambda
) \, \fet \chi (p) \, \quad \quad \text{with} \quad \fet \chi (p) =
(p, p^3 )^T\,. 
\ee
Here and in what follows, symbols in bold refer to matrix-valued
functions. 
In particular, $\fet \omega$ is a real $2 \times 2$ matrix
that depends on the cutoff $\Lambda$ and the on-shell momentum $k$. This is the type of 
potential we used to compute the LO scattering amplitude for  resonant
P-wave systems in section \ref{Halo_Nuclei},  cf.~Eq.~(\ref{BarePot}).

We consider a generic bare potential as defined in
Eq.~(\ref{PotBareRank2}), which
is required to yield a cutoff-independent off-shell scattering amplitude and
thus fulfills Eq.~(\ref{PreRG}). Following Ref.~\cite{Birse:2015iea}, we  
derive nontrivial fixed-point solutions of the RG equation (\ref{RGEq})
for the rescaled potential $\hat V (\hat p', \hat p, \hat k)$. We
start with rewriting Eq.~(\ref{RGEq}) in the form of the matrix
equation for $\hat {\fet \omega} (\hat k, \Lambda )$ defined via
$\hat
V (\hat p', \hat p, \hat k, \Lambda) =: \fet \chi^T (\hat p') \, \hat {\fet \omega}
(\hat k, \Lambda ) \, \fet \chi (\hat p)$: 
\be
\Lambda \frac{\partial \hat {\fet \omega}}{\partial \Lambda} =
\left( \begin{array}{cc} 1 & 0 \\ 0 & 3 \end{array} \right) \hat {\fet
  \omega}
+ \hat {\fet \omega} \left( \begin{array}{cc} 1 & 0 \\ 0 &
                                                           3 \end{array} \right)
                                                       + \hat k
                                                       \frac{\partial
                                                         \hat {\fet
                                                           \omega}}{
                                                         \partial \hat
                                                         k}
                                                       +  \hat {\fet
                                                         \omega}
                                                       +  \hat {\fet
                                                         \omega}
                                                       \frac{ 
                                                         {\fet \chi}
                                                         (1) \,  
                                                         {\fet \chi}^T
                                                         (1)  }{1-\hat
                                                         k^2 } \hat {\fet \omega}\,.
\ee
For invertible matrices $\hat {\fet \omega}$, the above RG equation
can be rewritten into a linear differential equation for $\hat {\fet \omega}^{-1}$: 
\be
- \Lambda \frac{\partial \hat {\fet \omega}^{-1}}{\partial \Lambda} =
\hat {\fet
  \omega}^{-1}\left( \begin{array}{cc} 1 & 0 \\ 0 & 3 \end{array} \right) 
+ \left( \begin{array}{cc} 1 & 0 \\ 0 &
                                                           3 \end{array}
                                                       \right) \hat {\fet \omega}^{-1}
                                                       - \hat k
                                                       \frac{\partial
                                                         \hat {\fet
                                                           \omega}^{-1}}{
                                                         \partial \hat
                                                         k}
                                                       +  \hat {\fet
                                                         \omega}^{-1}
                                                       +  
                                                       \frac{ 
                                                         {\fet \chi}
                                                         (1) \,  
                                                         {\fet \chi}^T
                                                         (1)  }{1-\hat
                                                         k^2 }\,,
\ee
which reduces to the uncoupled first-order partial differential
equations for the components of the matrix $\hat {\fet \omega}^{-1}$:
\bea
- \Lambda \frac{\partial \hat \omega_{11}^{-1}}{\partial \Lambda} &=&
3  \hat \omega_{11}^{-1} - \hat k \frac{\partial \hat
  \omega_{11}^{-1}}{\partial \hat k} + \frac{1}{1- \hat k^2}\,,
\nonumber \\
- \Lambda \frac{\partial \hat \omega_{12}^{-1}}{\partial \Lambda} &=&
5  \hat \omega_{12}^{-1} - \hat k \frac{\partial \hat
  \omega_{12}^{-1}}{\partial \hat k} + \frac{1}{1- \hat k^2}\,,
\nonumber \\
- \Lambda \frac{\partial \hat \omega_{22}^{-1}}{\partial \Lambda} &=&
7  \hat \omega_{22}^{-1} - \hat k \frac{\partial \hat
  \omega_{22}^{-1}}{\partial \hat k} + \frac{1}{1- \hat k^2}\,.
\eea
Here, we restrict ourselves to Hermitian potentials, so that
$\hat\omega_{12}^{-1} = \hat \omega_{21}^{-1} $. For $\Lambda$-independent
$\hat {\fet \omega}$, we can easily integrate these equations, subject to the
boundary condition that $\hat {\omega}^{-1}_{ij} (\hat k)$ are analytic functions of $\hat k^2$,
to obtain the potential corresponding to the rank-two fixed-point solution of the RG
equation
\bea
\label{Rank2FixedPoint}
\hat V_{\rm rank-2} (\hat p', \hat p, \hat k )&=&  \left( \hat p', \hat p'^3 \right) \left( \begin{array}{cc}
       {\rm Re }\, \hat J_3 (\hat k) &          {\rm Re }\, \hat J_5 (\hat k) \\[4pt]
      {\rm Re }\, \hat J_5 (\hat k) &          {\rm Re }\, \hat J_7 (\hat k)                                                       
\end{array}
\right)^{-1}
\left( \begin{array}{c} \hat p \\[4pt] \hat p^3 \end{array}
\right) \nonumber \\
&=&
- \frac{5 \hat p' \hat
  p}{5 - 7
\hat k^2 } \left[ 7 \left( \hat k^2 - \hat p'^2 \right)  \left(
  \hat k^2 - \hat p^2 \right)
+ \frac{6 \left( 5 - 7 \hat p'^2 \right) \left( 5 - 7 \hat p^2
  \right)}{8 + 80 \hat k^2 - 210 \hat k^4 - 15 \hat k^3 \left( 5 - 7
    \hat k^2 \right) \ln \frac{1+ \hat k}{1 - \hat k}}\right]\,,
\eea
where the rescaled dimensionless integrals $\hat J_n (\hat k)$ are defined in terms
of the integrals $J_n (k)$ from Eq.~(\ref{IntJ}) via
\be
\hat J_n (\hat k ) = \frac{2 \pi^2}{m}  \frac{1}{\Lambda^n} J_n (\hat
k \Lambda)\,.
\ee
Notice that apart from $\hat V_{\rm rank-2} (\hat p', \hat p, \hat k )$ and the
trivial potential $\hat V_{\rm T} (\hat p', \hat p, \hat k ) = 0$, any
potential corresponding to a non-invertible matrix $\hat {\fet \omega} (\hat k)$
with $\det \hat {\fet \omega} (\hat k ) = 0$ also represents a
fixed-point solution of the RG equation (which corresponds to $K(k)=0$).
Interestingly, if one drops the second term in the squared
brackets of Eq.~(\ref{Rank2FixedPoint}), the resulting scale-free
potential corresponds to such a fixed-point solution of the RG
equation with a non-invertible matrix $\hat {\fet \omega} (\hat k )$.

The above rank-two separable fixed point has nine relevant
perturbations, which can be parametrized by some (dimensionful) quantities
$\alpha_1, \, \alpha_2, \, \ldots , \, \alpha_9$. The
resulting RG-invariant rescaled potential with resummed relevant
perturbations can be written in the form
\be
\label{PotFullRescaled}
\hat V (\hat p', \hat p, \hat k, \Lambda  )= \left( \hat p', \hat p'^3 \right) \left( \begin{array}{ccc}
     \frac{\alpha_1}{\Lambda^3}  + \frac{\alpha_2}{\Lambda} \hat k^2 +   {\rm Re }\, \hat J_3 (\hat k) &&     \frac{\alpha_3}{\Lambda^5}  + \frac{\alpha_4}{\Lambda^3} \hat k^2 +  \frac{\alpha_5}{\Lambda} \hat k^4 +        {\rm Re }\, \hat J_5 (\hat k) \\[6pt]
   \frac{\alpha_3}{\Lambda^5}  + \frac{\alpha_4}{\Lambda^3} \hat k^2 +  \frac{\alpha_5}{\Lambda} \hat k^4 +      {\rm Re }\, \hat J_5 (\hat k) &&     \frac{\alpha_6}{\Lambda^7}  + \frac{\alpha_7}{\Lambda^5} \hat k^2 +  \frac{\alpha_8}{\Lambda^3} \hat k^4 +  \frac{\alpha_9}{\Lambda} \hat k^6 +        {\rm Re }\, \hat J_7 (\hat k)                                                       
\end{array}
\right)^{-1}
\left( \begin{array}{c} \hat p \\[6pt] \hat p^3 \end{array}
\right)\,,
\ee
and corresponds to the on-shell $K$-matrix
\be
\label{KMatFP2}
\frac{2 \pi^2 k^2}{m \, K(k) } = \alpha_1 + \alpha_2 k^2 -\frac{\left\{\alpha _3+k^2 \left[-\alpha _1+\alpha _4+\left(\alpha _5-\alpha _2\right)
   k^2\right]\right\}{}^2}{\alpha _6+k^2 \left\{-2 \alpha _3+\alpha _7+k^2 \left[\alpha _1-2 \alpha
   _4+\alpha _8+\left(\alpha _2-2 \alpha _5+\alpha _9\right) k^2\right]\right\}}\,.
\ee

It is not hard to see that the RG-invariant bare potential corresponding to the
subtractively renormalized scattering amplitude considered in Sections~\ref{Halo_Nuclei}
and \ref{FPsHPWs}, i.e.~the potential given in Eq.~(\ref{BarePot}) with
$J_n^R$ being replaced by  $J_n^{R, \; \infty} \equiv  \lim_{\Lambda  \to \infty } J_n^R $,
represents a special case of Eq.~(\ref{PotFullRescaled})  with
\be
\label{MatchingAlpha}
\alpha_1 = \alpha_4 = \alpha_8 = \frac{\mu_3^3}{3}\,, \quad 
\alpha_2 = \alpha_5 = \alpha_9= \mu_1 \,, \quad 
\alpha_7 = \frac{\mu_5^5}{5}\,, \quad 
\alpha_3 = \frac{1}{5 \tilde C_4^R } + \frac{\mu_5^5}{5}\,, \quad
\alpha_6 = - \frac{\tilde C_2^R}{5 (\tilde C_4^R )^2} + \frac{\mu_7^7}{7}\,.
\ee
The last two equalities show that the ``couplings'' $\mu_5$ and $\mu_7$
are indeed redundant as already pointed out in Section~\ref{FPsHPWs}.
Alternatively, following the procedure of that section, one can choose
to parametrize the theory directly in terms of the physical parameters $a^{-1}$, $r$
instead of the LECs $C_2^R$, $C_4^R$. For $\mu_5=0$,
Eq.~(\ref{MatchingAlpha}) then turns to\footnote{Notice that it is not
  always possible to express real values of the LECs $C_2^R$ and
  $C_4^R$ in terms of $a$ and $r$ if $\mu_5 \neq 0$.}
\be
\label{MatchingAlpha2}
\alpha_1 = \alpha_4 = \alpha_8 = \frac{\mu_3^3}{3}\,, \quad 
\alpha_2 = \alpha_5 = \alpha_9= \mu_1 \,, \quad 
\alpha_7 =0 \,, \quad 
\alpha_3 = \frac{2 (3 \pi a^{-1} - 2  \mu_3^3 )^2}{9 (\pi r + 4
  \mu_1)}\,, \quad 
\alpha_6 =-\frac{8 (3 \pi a^{-1} - 2  \mu_3^3 )^3}{27  (\pi r + 4
  \mu_1)^2}\,,
\ee
and the resulting effective range function coincides with
that given in Eq.~(\ref{ERFRen}). The scale-free limit of the
potential specified through the above equation is not uniquely
defined and depends on the order the limits $a^{-1} \to 0$, $r \to 0$,
$\mu_1 \to 0$ and $\mu_3 \to 0$ are taken. 
It corresponds to either $\hat V_{\rm U} (\hat p', \hat p, \hat k)$ in
Eq.~(\ref{UnitaryFP}) if one takes the limits e.g.~in the order
$\mu_3 \to 0$, $\mu_1 \to 0$, $r \to 0$ and $a^{-1} \to 0$ or
to $\hat V_{\rm rank-2} (\hat p', \hat p, \hat k)$
if the limits are taken e.g.~in the order 
$\mu_3 \to 0$, $\mu_1 \to 0$, $a^{-1} \to 0$ and $r \to 0$.

Interestingly, the above choice of the LECs  $C_2^R$ and
$C_4^R$ leading to Eq.~(\ref{MatchingAlpha2}) is not the only
possibility compatible with the given values of the scattering length and effective
range. One can see from Eq.~(\ref{KMatFP2}) that setting $\alpha_3 = 0$  via
$\tilde C_4^R = -1/\mu_5^5$ reproduces the first two terms in the ERE if
the subtraction scales $\mu_1$ and $\mu_3$ are tuned to the values
$\mu_1 = - \pi r/4$, $\mu_3^3 = 3 \pi a^{-1}/2$. The resulting potential
is determined by the parameters $\alpha_6$,  $\alpha_7$,  $a^{-1}$ and
$r$
via 
\be
\label{MatchingAlpha3}
\alpha_1 = \alpha_4 = \alpha_8 = \frac{3 \pi}{6} a^{-1}\,, \quad 
\alpha_2 = \alpha_5 = \alpha_9= - \frac{\pi r}{4} \,, \quad 
\alpha_3 = 0 \,.
\ee
Clearly, this solution is unphysical from the EFT point of view, 
since the reproduction of the scattering
length and effective range is achieved via fine tuning of
an infinite string of higher-order interactions, realized through a
particular choice of the subtraction scales $\mu_1$ and $\mu_3$. 

Specifying the theory by fixing the parameters $\alpha_i$ at some high resolution scale
$\Lambda$, one can follow the renormalized trajectories out of the
fixed point by considering
the RG flow down to $\Lambda \to 0$, which generally (but not
necessarily, see e.g.~Ref.~\cite{Birse:2015iea}) ends in the trivial
fixed point. Most importantly, regardless of a particular model, all
potentials that describe the systems we are interested in with the
coefficients in the ERE scaling according to Eqs.~(\ref{pc1}) and
(\ref{pars})\footnote{For the model in Eq.~(\ref{MatchingAlpha2}), this requires
  choosing $\mu_3\sim M_{\rm hi}$ or higher  to prevent the appearance of
a low-lying amplitude zero, while there are no
  restrictions on the choice of $\mu_1$. For the model in
  Eq.~(\ref{MatchingAlpha3}), the remaining coefficients $\alpha_6$, $\alpha_7$ can
  be chosen to scale with either powers of $M_{\rm lo}$ or $M_{\rm hi}$.}
feature a universal RG flow behavior once 
the cutoff is lowered below the hard scale $M_{\rm hi}$.
In particular, for the doubly fine-tuned scenario in Eq.~(\ref{pc1}),
they first get attracted to the unitary fixed point $\hat V_{\rm U}$
once $M_{\rm lo} \ll \Lambda \ll M_{\rm hi}$ before flowing towards
the trivial fixed point for $\Lambda \ll M_{\rm lo}$.  For other
systems, the running may be more exotic. For example, choosing
$\mu_3 \sim M_{\rm lo}$ in Eq.~(\ref{MatchingAlpha2}) while keeping the scaling
of $a^{-1}$ and $r$ as in Eq.~(\ref{pc1}), the potential gets
attracted to the fixed point  with a
non-invertible matrix $\hat {\fet \omega} (\hat k )$ corresponding to
the first term in the
square brackets of Eq.~(\ref{Rank2FixedPoint}) for $\Lambda \gtrsim
M_{\rm lo}$,  before finally 
running to the trivial fixed point for $\Lambda \ll M_{\rm lo}$.

\section{RG trajectory of a toy model potential with a long-range interaction}
\label{ToyModelRG}

As already pointed out in Section~\ref{FPsHPWs}, the Wilsonian RG analysis
does not provide a systematic expansion for the bare potential in the UV
regime with arbitrarily large cutoffs as relevant for the lcRG-invariant
approach. To illustrate this point, we consider below the  exact RG trajectory of
the toy model potential of Ref.~\cite{Epelbaum:2018zli} and compare it
to the LO cutoff-dependent potential of the lcRG-invariant approach.  
The toy model potential is given by
\be
V(r) = \frac{\alpha  \left(e^{- m_1 r}-e^{-M
   r}\right)}{r^3}+\frac{\alpha  \left(m_1-M\right) e^{- m_1 r}}{r^2}
   +\frac{\alpha  \left(M-m_1\right)^2 e^{-m_2 r}}{2 r} 
   - \frac{\alpha e^{- m_1 r}}{6}  \left(2 m_1-3
   m_2+M\right) \left(M-m_1\right)^2~,
\label{potentialdef}
\ee
where $M \sim {M_{\rm lo}}$ and $m_1 \sim m_2 \sim M_{\rm hi}$ refer
to the masses of the exchanged ``mesons''. For the demonstration
purpose below, we choose the numerical values of the parameters to be
$\alpha=5\times 10^{-5}\; {\rm MeV^{-2}}$, $M=138.5$~MeV, $m_1=750$~MeV and 
$m_2=1150$~MeV. Since the strength of the interaction $\alpha$ is 
equal for all terms, the potential $V(r)$ vanishes for $r\to 0$ but
behaves as $-\alpha\, e^{-M r}/r^3$ for large $r$. More details of the model can be found in
Ref.~\cite{Epelbaum:2018zli}.
Regarding the above potential as an "underlying" interaction model, we
construct below 
the LO EFT approximation using the lcRG-invariant approach and compare
it to the exact Wilsonian RG trajectory of the "underlying" potential.  

\begin{figure}[t]
\includegraphics[width=0.4\textwidth]{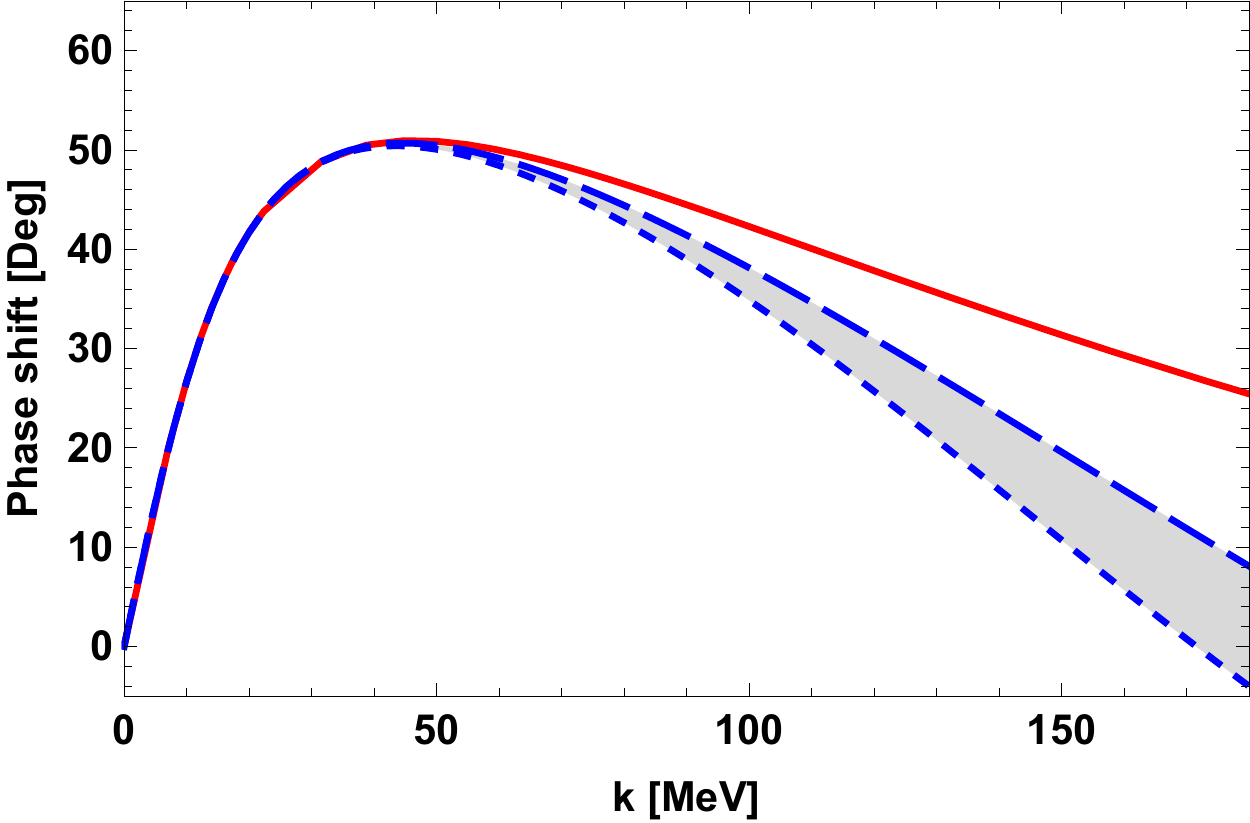}
\caption{The S-wave phase shift for the underlying toy model and the
  LO approximation shown by the solid red and dashed blue lines,
  respectively.  
  The short-  and long-dashed lines 
  correspond to the cutoff values of $300$ and $450$~MeV,
  respectively, while the
  shaded area visualizes the corresponding cutoff dependence of the
  phase shift. }
\label{PhShs} 
\end{figure}

We consider the LS equation for the S-wave $K$-matrix in the center-of-mass frame of two particles 
with equal masses $m=1000$~MeV 
\be
K(p',p, k) =
V (p ',p)+  m \dashint_0^\infty\frac{d l\, l^2}{2\pi^2}
\, V(p',l) \,\frac{ 1}{k^2-l^2}\, K(l,p, k) \, .
\label{tEq}
\ee
The resulting phase shift as a function of the momentum $k$ is shown
by the red line in Fig.~\ref{PhShs}. 
At low energies, we can integrate out the high-energy modes and obtain
the scattering amplitude by solving the regularized equation 
\be
K(p',p, k) =  V (p ',p, k, \Lambda ) +  m \dashint_0^\Lambda \frac{d
  l\, l^2}{2\pi^2}
\, V(p',l, k, \Lambda ) \, \frac{ 1 }{k^2-l^2} \, K(l,p, k)\,,
\label{tEqC}
\ee
where  the potential $ V (p ',p, k, \Lambda ) $ satisfies the equation 
\begin{equation}
V(p',l, k, \Lambda ) =  V (p ',p)  +  m \dashint_\Lambda^\infty
\frac{d l\, l^2}{2\pi^2} \,
 V (p ',l) 
\,\frac{1}{k^2-l^2}\, V(l,p, k, \Lambda ) \,.
\label{vEqC}
\end{equation}
Equation~(\ref{vEqC}) gives the exact Wilsonian RG trajectory of the effective potential.

For low cutoffs, the potential $V(p',l, k, \Lambda )$ can be
approximated by $V_{\rm LO}$,  the Fourier transform of the
delta-potential plus the long-range part of the interaction $ -{\alpha\, e^{-Mr}}/{r^3}$.
Choosing some cutoff value between small and large
scales of the problem ($\Lambda \sim 0.4 $~GeV), we adjust the
strength of the contact interaction $C(\Lambda)$ 
such that at very low energies, the phase shifts of the underlying
model are well described by the solution to the equation:
\begin{equation}
K_{\rm LO}(p',p, k)  =   V_{\rm LO} (p ',p) +  m \dashint_0^\Lambda \frac{d
  l\, l^2}{2\pi^2}
\, V_{\rm LO }(p',l ) \, \frac{ 1 }{k^2-l^2} \, K_{\rm LO }(l,p, k)\,.
\end{equation}
The resulting phase shifts are plotted as a function of $k$ in
Fig.~\ref{PhShs} together with the phase shifts corresponding to the
underlying model. 

\begin{figure}[t]
\includegraphics[width=0.8\textwidth]{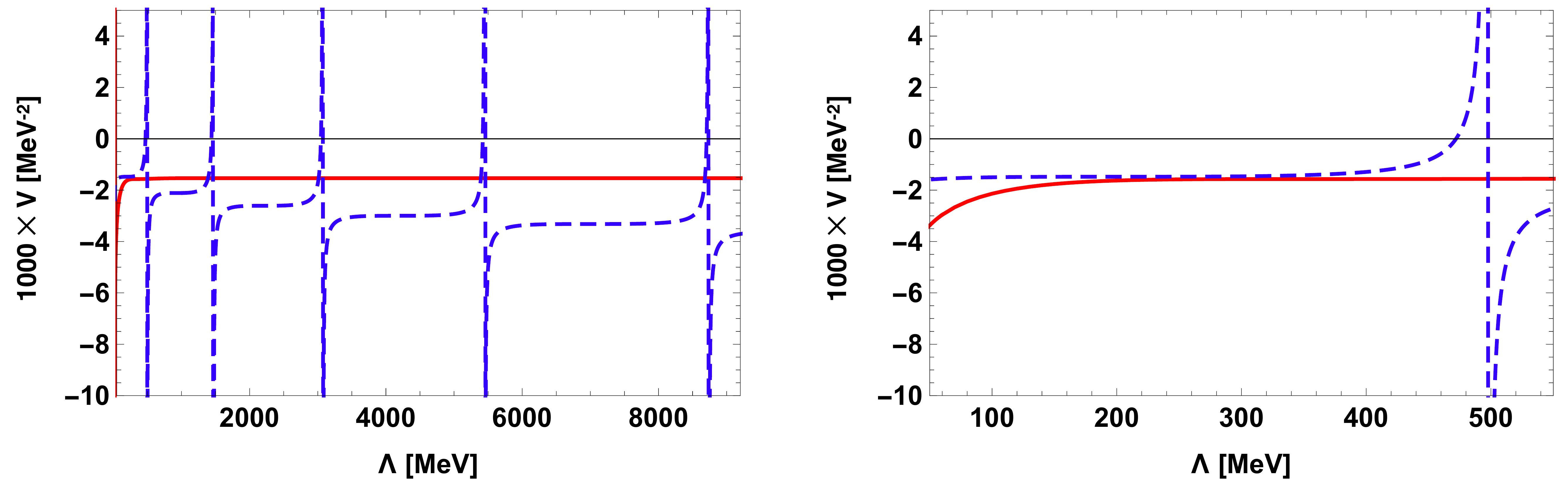}
\caption{RG trajectories
of the on-shell potentials ($k =20$~MeV) for the toy model. Solid
(red)  and the dashed (blue) lines correspond to the underlying toy
model and the LO approximation, respectively, as discussed in the text. 
  The right panel is a zoomed version of the left one showing the
  low-cutoff region. }
\label{RGTrajectories} 
\end{figure}

Following the lcRG-invariant approach, we now take arbitrarily large values of
the cutoff in the LO approximation and obtain, by adjusting the
contact interaction as a function of $\Lambda$, (almost)
cutoff-independent results for phase shifts at low energies.
The corresponding cutoff-dependent on-shell potential for $k=20$~MeV
is plotted  in Fig.~\ref{RGTrajectories} together with the exact RG
trajectory of the underlying toy-model potential, obtained by solving
numerically Eq.~(\ref{vEqC}).  
While the LO potential does approximate well the exact RG trajectory
for $\Lambda$ around $\sim 300$~MeV,  the limit-cycle behavior of the
LO potential for larger values of the cutoff is just an artifact of the
lcRG-invariant approach.  

\end{document}